\documentclass[%
aip,
jcp,
amsmath,amssymb,
preprint,%
]{revtex4-2}

\usepackage{graphicx}
\usepackage{dcolumn}
\usepackage{bm}

\usepackage[utf8]{inputenc}
\usepackage[T1]{fontenc}
\usepackage{mathptmx}
\usepackage{etoolbox}

\usepackage{comment}

\usepackage{enumerate}
\usepackage{algorithm}
\usepackage{algorithmicx}
\usepackage{algpseudocode}
\usepackage{paralist}
\usepackage{threeparttable}
\usepackage{threeparttablex}
\usepackage{booktabs}
\usepackage{makecell}
\usepackage{multirow}
\usepackage{color}

\algnewcommand{\Inputs}[1]{
  \State \textbf{Inputs:}
  \Statex \hspace*{\algorithmicindent}\parbox[t]{.8\linewidth}{\raggedright #1}
}
\algnewcommand{\Initialize}[1]{
  \State \textbf{Initialize:}
  \Statex \hspace*{\algorithmicindent}\parbox[t]{.8\linewidth}{\raggedright #1}
}
\algnewcommand{\Outputs}[1]{
  \State \textbf{Outputs:}
  \Statex \hspace*{\algorithmicindent}\parbox[t]{.8\linewidth}{\raggedright #1}
}
\algnewcommand\algorithmicswitch{\textbf{switch}}
\algnewcommand\algorithmiccase{\textbf{case}}
\algdef{SE}[SWITCH]{Switch}{EndSwitch}[1]{\algorithmicswitch\ #1\ \algorithmicdo}{\algorithmicend\ \algorithmicswitch}%
\algdef{SE}[CASE]{Case}{EndCase}[1]{\algorithmiccase\ #1}{\algorithmicend\ \algorithmiccase}%
\algtext*{EndSwitch}%
\algtext*{EndCase}%

\newfloat{Algorithm}{htbp}{alg}
\floatname{Algorithm}{Algorithm}

\makeatletter
\def\@email#1#2{%
 \endgroup
 \patchcmd{\titleblock@produce}
  {\frontmatter@RRAPformat}
  {\frontmatter@RRAPformat{\produce@RRAP{*#1\href{mailto:#2}{#2}}}\frontmatter@RRAPformat}
  {}{}
}%
\makeatother

\UseRawInputEncoding
\begin{document}

\preprint{}

\title[Approximating First Hitting Point Distribution in Milestoning for Rare Event Kinetics]{Approximating First Hitting Point Distribution in Milestoning for Rare Event Kinetics}
\author{Ru Wang}
\affiliation{
Qingdao Institute for Theoretical and Computational Sciences, Institute of Frontier and Interdisciplinary Science, Shandong University, Qingdao, Shandong 266237, P. R. China}

\author{Hao Wang*}
\email{wanghaosd@sdu.edu.cn} %
\affiliation{
Qingdao Institute for Theoretical and Computational Sciences, Institute of Frontier and Interdisciplinary Science, Shandong University, Qingdao, Shandong 266237, P. R. China}

\author{Wenjian Liu}
\affiliation{
Qingdao Institute for Theoretical and Computational Sciences, Institute of Frontier and Interdisciplinary Science, Shandong University, Qingdao, Shandong 266237, P. R. China}

\author{Ron Elber}
\affiliation{
Oden Institute for Computational Engineering and Sciences, University of Texas at Austin, Austin, Texas 78712, United States}
\affiliation{Department of Chemistry, University of Texas at Austin, Austin, Texas 78712, United States}


\begin{abstract}
Milestoning is an efficient method for rare event kinetics calculation using short trajectory parallelization. Mean first passage time (MFPT) is the key kinetic output of Milestoning, whose accuracy crucially depends the initial distribution of the short trajectory ensemble. The true initial distribution, i.e., first hitting point distribution (FHPD), has no analytic expression in the general case. Here, we introduce two algorithms, local passage time weighted Milestoning (LPT-M) and Bayesian inference Milestoning (BI-M), to accurately and efficiently approximate FHPD for systems at equilibrium condition. Starting from sampling Boltzmann distribution on milestones, we calculate the proper weighting factor for the short trajectory ensemble. The methods are tested on two model examples for illustration purpose. Both methods improve significantly over the widely used classical Milestoning method in terms of the accuracy of MFPT. In particular, BI-M covers the directional Milestoning method as a special case in the deterministic Hamiltonian dynamics. LPT-M is especially advantageous in terms of computational costs and robustness with respect to the increasing number of intermediate milestones. Furthermore, a locally iterative correction algorithm for non-equilibrium stationary FHPD is developed for exact MFPT calculation, which can be combined with LPT-M/BI-M and is much cheaper than the exact Milestoning method.   
\end{abstract}

\maketitle

\onecolumngrid

\section{Introduction}
Atomically detailed computer simulations are a useful tool to study thermodynamics and kinetics of molecular systems\cite{MolSimu02}. However, these calculations are expensive, and are not always feasible. The calculations of kinetics are particularly demanding since the samples are of complete reactive trajectories from reactants to products. This ensemble should be contrasted with the ensemble of configurations required for thermodynamic averages. Individual configurations are of orders of magnitude cheaper to sample than complete reactive trajectories.

Therefore, in the last two decades considerable efforts were invested to develop theories and algorithms for exact and approximate simulations of kinetics\cite{MDcondensed}. An attractive approach is based on the division of phase space into partitions or cells. We consider local kinetics between cells and the aggregation of the local kinetic results to compute the flux through the entire reaction space. Methods like Transition Interface Sampling\cite{TIS03,TIS05} (TIS), Forward Flux Sampling\cite{FFlux06,FFlux09} (FFS), Non-equilibrium Umbrella Sampling\cite{NEUS09} (NEUS) and Milestoning\cite{CM04} exploit the events of crossing boundaries between cells to estimate kinetic observables such as Mean First Passage Time (MFPT). We call the boundaries between cells milestones. Other approaches like Markov State Model\cite{MSM11,MSM14,MSM18} (MSM) or the Weighted Ensemble\cite{WE96,WE10} (WE) approach, consider changes in the populations in the cells (not necessarily crossing the boundaries). 

Milestoning provides a general mathematical framework for finite-state and continuous-time modeling of rare event kinetics. The accuracy of Milestoning crucially depends on two factors: (i) the initial distribution of short trajectory ensemble; (ii) statistical adequacy of short trajectory sampling. The first determines the systematic error, while the second determines the statistical error. Many variants\cite{MarkovM09,Tilting09,DiM10,ExM15,WARM18,WEM20,MileW20} have been developed to address these two issues since the first introduction of classical Milestoning\cite{CM04} (CM).

In this paper we consider the crossing point distribution on the milestone. We consider an exact definition and approximations. Each crossing point (a phase space configuration) has a weight of one for the long trajectory sampling. As such, the weight is nonuniform for sampling of a trajectory crossing. However, the entire weight of a trajectory including all crossing points must be one in the calculations of fluxes and kinetics. 
For a trajectory with multiple crossings, there are many ways of assigning weights to each crossing point. We may assign a weight of one for the first and zeros to the rest of crossing points. This choice leads to a distribution called the first hitting point distribution (FHPD). We denote it by $f(x)$ where $x$ is a phase space crossing point. The FHPD was discussed extensively in the context of Transition Path Theory\cite{TPT06,TPT10} (TPT) and was shown to provide the exact MFPT within the Milestoning theory\cite{AssuMile08}. FHPD has no analytic expression in the general case and needs to be approximated numerically.

In CM Boltzmann distribution (BD) constrained to a milestone hypersurface is used as a rough approximation of FHPD in the canonical ensemble. When the spacing between milestones is large and the potential energy surface is not steep, trajectories initiated from a milestone have sufficient time to relax to a local equilibrium before hitting a different one. As such, the BD approximation is sound. Using solvated alanine dipeptide as an example, CM was illustrated to work well if the velocity correlation function decays to zero before hitting a new milestone\cite{Elber07}. However, for highly activated processes, milestones need to be placed closer, otherwise hitting events uphill in energy are challenging to sample. In this case, the discrepancy between BD and FHPD is appreciable. In directional Milestoning\cite{DiM10} (DiM), FHPD is obtained from a resampling procedure from an initial BD constrained to the milestone by removing those samples that are not real first hitting points. 
The selection is done by running trajectories from sampled configurations backward in time and checking if other re-crossing events are found before the trajectories hit another milestone. 
This selection rule was proposed for trajectory evolution following the deterministic Hamiltonian dynamics. 
However, the removal of re-crossing trajectories results in considerable loss of statistics. In exact Milestoning\cite{ExM15} (ExM), FHPD is iteratively updated using BD as an initial guess. Under mild conditions, the iterations converge the distribution on milestones to FHPD\cite{MathExM16}. However, the iterations are usually time consuming and more efficient approaches are desired.

The statistical adequacy of trajectory sampling is a trickier issue. Nonetheless, an infinite MFPT output resulting from a disconnected Milestoning network clearly indicates an insufficient trajectory sampling. One possible scenario for this is the rare hitting events uphill in energy in an activated process. The biasing-and-reweighting method for trajectory sampling can be utilized to significantly enhance those otherwise rare transitions\cite{WARM18,MileW20}.

In this paper, we develop two algorithms of approximating equilibrium FHPD. One is based on local passage time of a trajectory crossing a milestone. All configurations sampled on a milestone are retained. As a result, no statistics is lost. The other is based on Bayesian inference, which generalizes DiM to stochastic dynamics. Furthermore, a locally iterative correction algorithm for non-equilibrium stationary FHPD is developed for exact MFPT calculation. 

The remainder of this paper is organized as follows. First in Sec. \ref{Milestoning} we briefly review the Milestoning framework emphasizing on the effect of FHPD on MFPT calculation. Next, in Sec. \ref{Long Traj} we show how Milestoning can be combined with a long trajectory simulation to serve as the MFPT reference. In Sec. \ref{FHPD approx}, we show the details of the algorithms. Finally, in Sec. \ref{Results} we illustrate the performance of our methods on two model examples.

\section{Methods}\label{Methods}
\subsection{Milestoning Backdrop}\label{Milestoning}
We here only summarize the essential aspects of the Milestoning algorithm with the focus on illustrating how the FHPD affects the MFPT computation. Readers are referred to recent reviews for more detailed discussions of Milestoning algorithms\cite{RevMile20,RevMile21} and software implementation\cite{ScMiles2}.

\begin{figure}[h]
\centering
\begin{tabular}{cc}
\includegraphics[height=7cm]{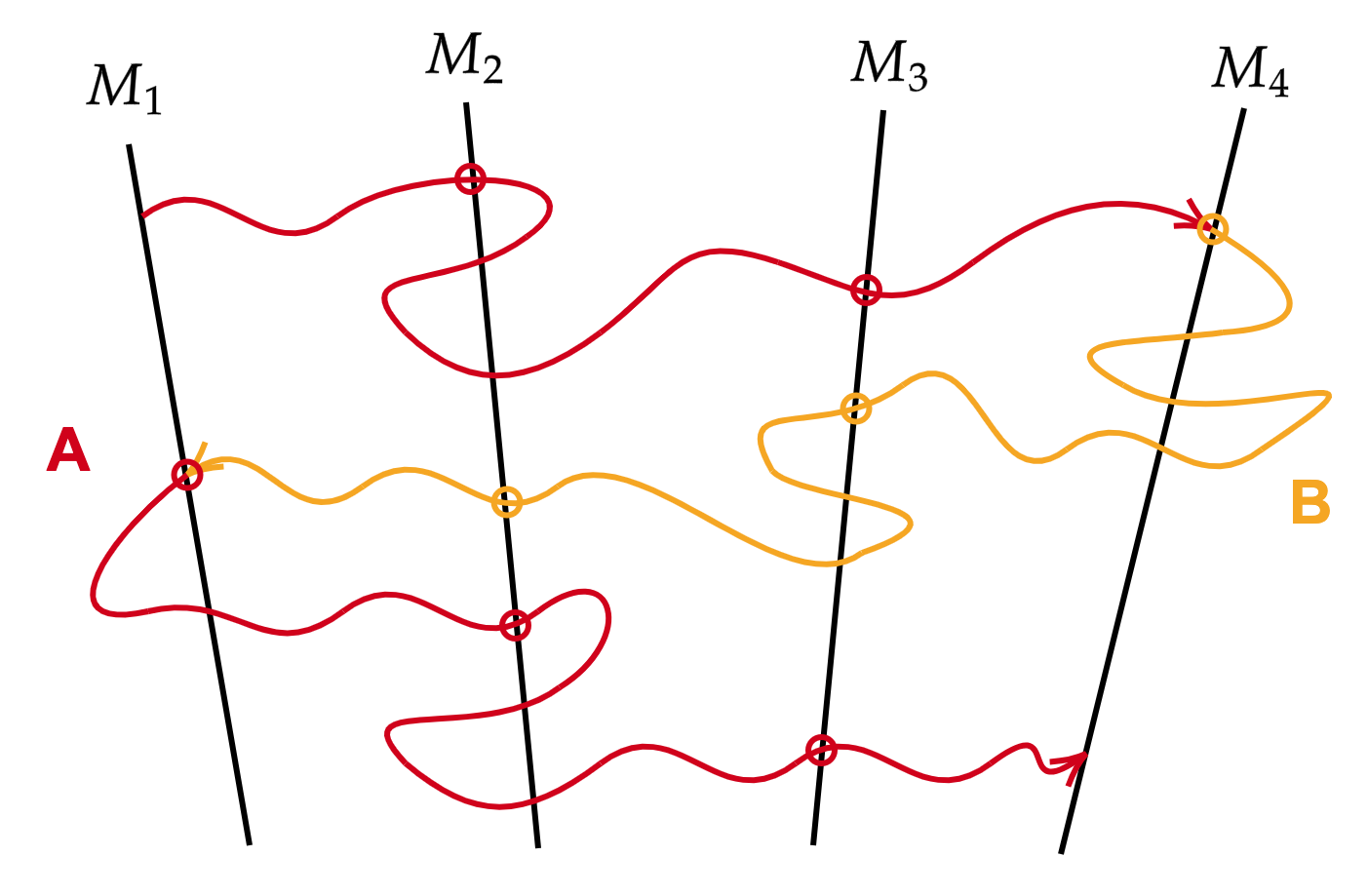}\\
(a)\\
\includegraphics[height=7cm]{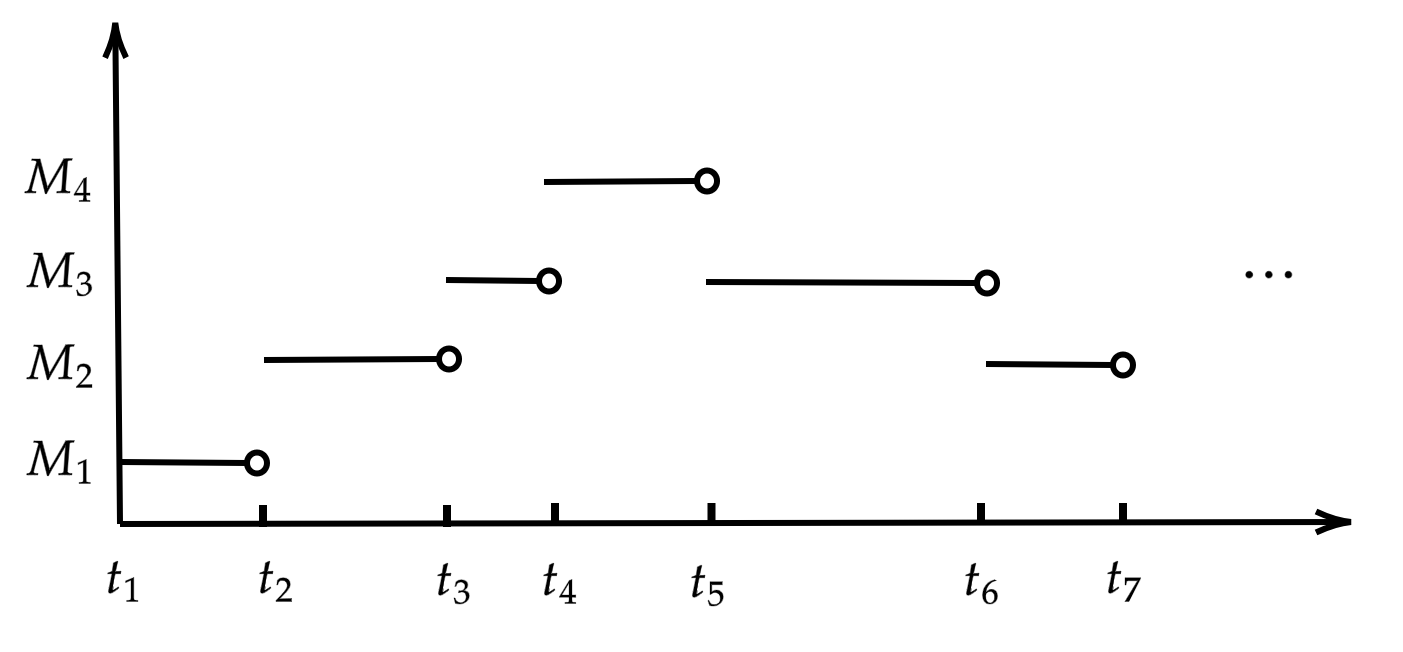}\\
(b)
\end{tabular}
\caption{(a) Schematic diagram of an equilibrium long trajectory transiting between $A$ (reactant) and $B$ (product). First hitting points are marked as circles. (b) The path history mapped into the discretized milestone state space.}\label{fig_sketch}
\end{figure}

Consider a phase space of dimension $\mathbb{R}^{6N}$ with $N$ denoting the number of atoms. We are interested in computing the MFPT from a metastable region $A$ (the reactant state) to another disjoint metastable region $B$ (the product state). When the transition from $A$ to $B$ is an activated process or dominated by slow diffusion on a rugged energy landscape, a straightforward molecular dynamics (MD) simulation is usually infeasible. 
However, to understand the key elements underlying Milestoning, it is useful to consider an infinitely long trajectory transiting back and forth between $A$ and $B$ (cf. Fig. \ref{fig_sketch} (a)). Based on which state ($A$ or $B$) the trajectory last visits, the path ensemble can be divided into two classes: $A\rightarrow B$ and $B\rightarrow A$. 

In Milestoning, the configuration space is partitioned into small compartments, and the interfaces between compartments are called milestones, which are denoted by $\{M_1,M_2,...,M_n\}$. The current state of the long trajectory is determined by the last milestone it crossed. As such, the path history in high-dimension phase space is mapped into a discretized milestone state space (Fig. \ref{fig_sketch} (b)), from which the transition probability $K_{\alpha\beta}, \alpha,\beta\in\{M_1,M_2,...,M_n\}$, between nearby milestones and the mean dwelling time $t_\alpha$ on each milestone can be calculated. 
It has been shown that MFPT from $A$ to $B$ can be exactly calculated once $\{K_{\alpha\beta}\}$ and $\{t_\alpha\}$ of the partial path ensemble $A\rightarrow B$ are known\cite{AssuMile08,ExM15}. 

The goal of Milestoning is to calculate $\{K_{\alpha\beta}\}$ and $\{t_\alpha\}$ from short trajectory simulations, which are directly initiated from milestones. These short trajectories continue until they hit a different milestone for the first time.
Short trajectories initiated from different milestones can run in trivial parallelization. Local transition probabilities and time are finally integrated together by solving a matrix equation (Eqs. \eqref{MFPT absorb} or \eqref{MFPT cyc}). This is why Milestoning is much more efficient than a brute-force long trajectory simulation of $A\rightarrow B$ transitions. 
To be consistent with the direct long trajectory simulation of $A\rightarrow B$ transitions, the initial distribution of the short trajectory ensemble should be the non-equilibrium stationary FHPD of the partial path ensemble $A\rightarrow B$ (red circles in Fig. \ref{fig_sketch} (a)). In contrast, the equilibrium FHPD consists of contributions from both path ensemble $A\rightarrow B$ and $B\rightarrow A$ (red and yellow circles in Fig. \ref{fig_sketch} (a)).

The non-equilibrium stationary FHPD of the partial path ensemble $A\rightarrow B$ has no analytic expression in general case. Therefore it needs to be approximated numerically. In CM\cite{CM04} it is roughly approximated by BD, while in ExM\cite{ExM15} it is iteratively corrected on the basis of CM by solving the self-consistent equation,
\begin{equation}
f^{(i+1)}_\beta(x_\beta) =\frac{1}{q^{(i)}_\beta}\sum_{\alpha\neq\beta}q^{(i)}_\alpha K^{(i)}_{\alpha\beta}(x_\beta),
\label{FHPD iter}
\end{equation}
where $f^{(i)}_\alpha(x_\alpha)$ ($x_\alpha=(r^{3N}_\alpha,p^{3N}_\alpha)$ denotes a phase space configuration on the milestone $\alpha$) is the approximate non-equilibrium stationary FHPD of the $i$th iteration, $q_\alpha^{(i)}$ is the stationary flux through milestone $\alpha$, and $K^{(i)}_{\alpha\beta}(x_\beta)\equiv\int_\alpha dx_\alpha\int dt\cdot f_\alpha^{(i)}(x_\alpha)K(x_\alpha,x_\beta,t)$ is averaged transition probability from milestone $\alpha$ to $x_\beta$.
A full iteration loop involves four sequential steps: (i) generate initial phase space configurations on milestones according to $f^{(i)}_\beta(x_\beta)$; (ii) Run trajectories forward in time from sampled configurations until they hit a different milestone for the first time; (iii) Solve Eq. \eqref{flux cyc} to obtain stationary flux $q_\beta^{(i)}$; (iv) Resample first hitting points on milestones according to their contributions. Suppose we want to sample $n_\beta$ points on a milestone $\beta$, which constitutes a sampled version of $f_\beta^{(i+1)}(x_\beta)$. For each milestone $\alpha$ that is directly connected to $\beta$, we need to uniformly sample $n_\beta q_\alpha^{(i)}K_{\alpha\beta}^{(i)}/q_\beta^{(i)}$ times from first hitting points coming from milestone $\alpha$, where $K_{\alpha\beta}^{(i)}$ is the transition probability from $\alpha$ to $\beta$ as defined in Eq. \eqref{K matrix} in the $i$th iteration. Readers are referred to the Ref. \cite{ExM15} for more details of the ExM algorithm.

Given an approximation of the non-equilibrium stationary FHPD of the partial path ensemble $A\rightarrow B$ on milestone $\alpha$, $f_\alpha(x_\alpha)$, the transition probability $K_{\alpha\beta}$ and the mean dwelling time $t_\alpha$ are calculated by\cite{ExM15}
\begin{equation}
K_{\alpha\beta} = \int_\alpha dx_\alpha\int_\beta dx_\beta\int dt \cdot f_\alpha({x_\alpha}) K(x_\alpha,x_\beta,t),
\label{K matrix}
\end{equation}
\begin{equation}
t_\alpha = \sum_{\beta\neq \alpha}\int_\alpha dx_\alpha\int_\beta dx_\beta\int dt \cdot t\cdot f_\alpha(x_\alpha)K(x_\alpha,x_\beta,t),
\end{equation}
where $K(x_\alpha,x_\beta,t)$ is the transition kernel starting from $x_\alpha$ and ending at $x_\beta$ during a time interval $t$. The concrete form of $K(x_\alpha,x_\beta,t)$ depends on the equation of motion used to evolve the dynamics. In practice, $K_{\alpha\beta}=n_{\alpha\beta}/n_\alpha$ where $n_{\alpha\beta}$ out of $n_\alpha$ trajectories initiated on milestone $\alpha$ first hit milestone $\beta$, and $t_\alpha=\sum_{i=1}^{n_\alpha} t_\alpha(i)/n_\alpha$ where $t_\alpha(i)$ is the lifetime of the $i$th trajectory initiated on milestone $\alpha$. 

By the first-step analysis method, the MFPT, $\tau_\alpha$, from a certain milestone $\alpha$ to the predefined product milestone $\gamma$ satisfies a recurrence relation,
\begin{equation}
\tau_\alpha = t_\alpha + \sum_{\beta\neq\alpha} K_{\alpha\beta} \tau_\beta.\label{MFPT recurrence}
\end{equation}
It has a simple physical explanation that the $\alpha\rightarrow\gamma$ transition consists of a first step to a nearby milestone $\beta$ and then a final $\beta\rightarrow\gamma$ transition. Eq. \eqref{MFPT recurrence} can be written in a more compact matrix equation form,
\begin{equation}
(\mathbf{I}-\mathbf{K}_A)\mathbf{\tau}=\mathbf{t},\label{MFPT absorb}
\end{equation}
where $\mathbf{I}$ is an $n\times n$ identity matrix, $\mathbf{K}_A$ is the transition probability matrix with element $K_{\alpha\beta}$ except the row corresponding to the product state, and $\mathbf{\tau}$ and $\mathbf{t}$ are column vectors. By definition $\tau_\gamma=0$. The $\mathbf{K}_A$ matrix is imposed with absorbing boundary condition at the product state, i.e., $K_{A,\gamma\alpha}=0$ for all $\alpha\in\{M_1,M_2,...,M_n\}$, and $t_\gamma=0$ to be consistent. 
This amounts to removing trajectories out of the system once they arrive at the product state. 

Alternatively, cyclic boundary condition can be set at the product state in the transition probability matrix. Trajectories are re-injected into the reactant state $\xi$ once they arrive at the product state $\gamma$, i.e., $K_{C,\gamma\alpha}=\delta_{\alpha\xi}$ for all $\alpha\in\{M_1,M_2,...,M_n\}$ with $\delta_{\alpha\xi}$ being the Kronecker delta function. The non-equilibrium stationary flux $\{q_\alpha\}$ through each milestone obeys the eigenvalue equation
\begin{equation}  
\mathbf{q}^T=\mathbf{q}^T\mathbf{K}_C. \label{flux cyc}
\end{equation}
The MFPT $\tau_\xi$ is then calculated as 
\begin{equation}
\tau_\xi=\sum_{\alpha\neq\gamma} q_{\alpha} t_\alpha/q_\gamma, \label{MFPT cyc}
\end{equation}
where the summation is over all milestones except the product. The term $q_\alpha t_\alpha$ is the stationary probability of the last crossed milestone being $\alpha$ in the partial path ensemble $A\rightarrow B$. Therefore, Eq. \eqref{MFPT cyc} bears the meaning of "population over flux".

\subsection{Milestoning Combined with a Long Trajectory Simulation as Reference}\label{Long Traj}
As discussed above, the non-equilibrium stationary FHPD of the partial path ensemble $A\rightarrow B$ is required for exact MFPT calculation. 
A straightforward way to achieve this is by conducting an equilibrium long trajectory that transits back and forth between $A$ and $B$ and then selecting only the path ensemble $A\rightarrow B$ for analysis. 
However, when the reactant $A$ is a metastable state, there exists a more efficient method to simulate the path ensemble $A\rightarrow B$. 
In practice, the simulation proceeds as follows. Draw a set of configurations on the reactant milestone in the canonical ensemble. For each configuration, initial velocities are drawn from the Maxwell distribution and a trajectory is simulated forward in time until it reaches the product state $B$. Throughout this simulation, absorbing or cyclic boundary conditions are imposed at the product milestone $B$.

It is noteworthy that the ideally preferred initial distribution on the reactant milestone should be FHPD. However, when considering only two milestone states $A$ and $B$, the discrepancies between FHPD and BD for initial sampling at the reactant state is negligible. This is because we are focused on a rare transition process $A\rightarrow B$, where the time taken to reach $B$ is lengthy. Consequently, the trajectory has sufficient time to relax and erase its initial state memory before reaching $B$. For scenarios with more than two milestones, the BD approximation at the reactant needs to be carefully assessed. This can be achieved either by comparing the MFPT with the two-state result as the number of milestones increases, or by evaluating the reactant milestone's lifetime against the velocity decorrelation time. If the velocity correlation function diminishes before reaching a new milestone, the differences between FHPD and BD become negligible. This principle was demonstrated by Elber and co-workers in a study of solved alanine dipeptide\cite{Elber07}.

A long trajectory simulation in cyclic boundary condition (denoted as Long-C) naturally yields a transition probability matrix $\mathbf{K}_C$ of the same boundary condition. By solving Eqs. \eqref{flux cyc} and \eqref{MFPT cyc}, the exact MFPT can be obtained. 

Since our starting point of FHPD approximation in Sec. \ref{LPT} and \ref{BI} is the BD, which includes all crossing points on a milestone, the equilibrium FHPD is eventually derived. It encompasses contributions from both the path ensemble $A\rightarrow B$ and $B\rightarrow A$ ( represented by the red and yellow circles in Fig. \ref{fig_sketch} (a)). 
Therefore, we also conduct a single equilibrium long trajectory transiting back and forth between $A$ and $B$ for comparison. This simulation amounts to a reflecting boundary condition in the sense that the sum of transition probabilities from product milestone $\gamma
$ to its neighboring milestones is one. A transition is counted each time the long trajectory hits a milestone other than the one it was currently assigned. In ergodic systems, a natural reflecting boundary often exists at the product state. That implies that trajectories arriving at $B$ will ultimately return to $A$ without requiring manual velocity flipping. This equilibrium long trajectory simulation provides an unbiased estimation of the equilibrium FHPD. 

A single equilibrium long trajectory simulation (denoted as Long-R) automatically leads to a transition probability matrix $\mathbf{K}_R$ of the reflecting boundary condition. To calculate MFPT using Eq. \eqref{flux cyc}, the matrix $\mathbf{K}_R$ is first adjusted to cyclic boundary condition by setting $K_{R,\gamma\alpha}=\delta_{\alpha\xi}$ for all $\alpha$ with $\xi$ being the reactant milestone and $\gamma$ being the product milestone. Subsequently, the modified $\mathbf{K}_R$ is substituted into Eq. \eqref{flux cyc} in replace of $\mathbf{K}_C$ and finally Eq. \eqref{MFPT cyc} is solved to determine the MFPT. It is important to note that the MFPT calculated in this way is not theoretically exact due to the fact that the transition probability of intermediate milestones include contributions from both path ensemble $A\rightarrow B$ and $B\rightarrow A$. However, this approach serves as the accuracy limit for the two algorithms described in Sec. \ref{LPT} and \ref{BI}. The MFPT calculated in this way is shown to only slightly deviate from that obtained in cyclic boundary condition in Sec. \ref{Results}.

\subsection{Approximating FHPD}\label{FHPD approx}
The accurate estimation of the non-equilibrium stationary FHPD of the partial path ensemble $A\rightarrow B$ required for exact MFPT calculation can be computationally expensive. This is due to the necessity of retracing trajectories all the way back to either the reactant $A$ or the product $B$ to verify their origins. Instead, we choose to approximate the equilibrium FHPD, which encompasses contributions from both path ensembles $A\rightarrow B$ and $B\rightarrow A$. While this approach is not exact for MFPT calculations, the accuracy of predicted MFPT has been observed to improve significantly compared to CM. Furthermore, we introduce a locally iterative correction algorithm for exact MFPT calculation. This algorithm offers a more stable and efficient solution compared to the ExM method.

\subsubsection{Local Passage Time Weighting} \label{LPT}  
We start from BD sampling on each milestone in the canonical ensemble and then calculate the weighting factor needed for approximating equilibrium FHPD. 

We first consider unconstrained sampling of BD on a milestone in an equilibrium long trajectory simulation. In MD simulations, the time integration is discretized and configurations are saved every time step $\Delta t$. These configurations would not fall exactly on a milestone (a hypersurface of measure zero) defined by $M(r^{3N})=m$, where $M(r^{3N})$ is a coarse function of Cartesian coordinates, e.g., interatomic distances, torsion angles etc., used to partition the configuration space. Instead, we consider an infinitesimal interval $[m,m+dM]$. Whenever configurations fall into this interval, they are treated as falling on the milestone. The collection of configurations that fall within $[m,m+dM]$ constitutes BD on the milestone.

During the $l$th crossing event, the number of configurations $n_l$ that fall into the interval $[m,m+dM]$ is proportional to its local passage time $dt_l$, $n_l=dt_l/\Delta t$. Or equivalently, $n_l$ is proportional to $|v_l^\perp|^{-1}$ with $v_l^\perp$ denoting the component of velocity normal to the milestone of crossing since $dt_l=dM/|v_l^\perp|$.  

\begin{figure}[h]
\centering
\includegraphics[height=7cm]{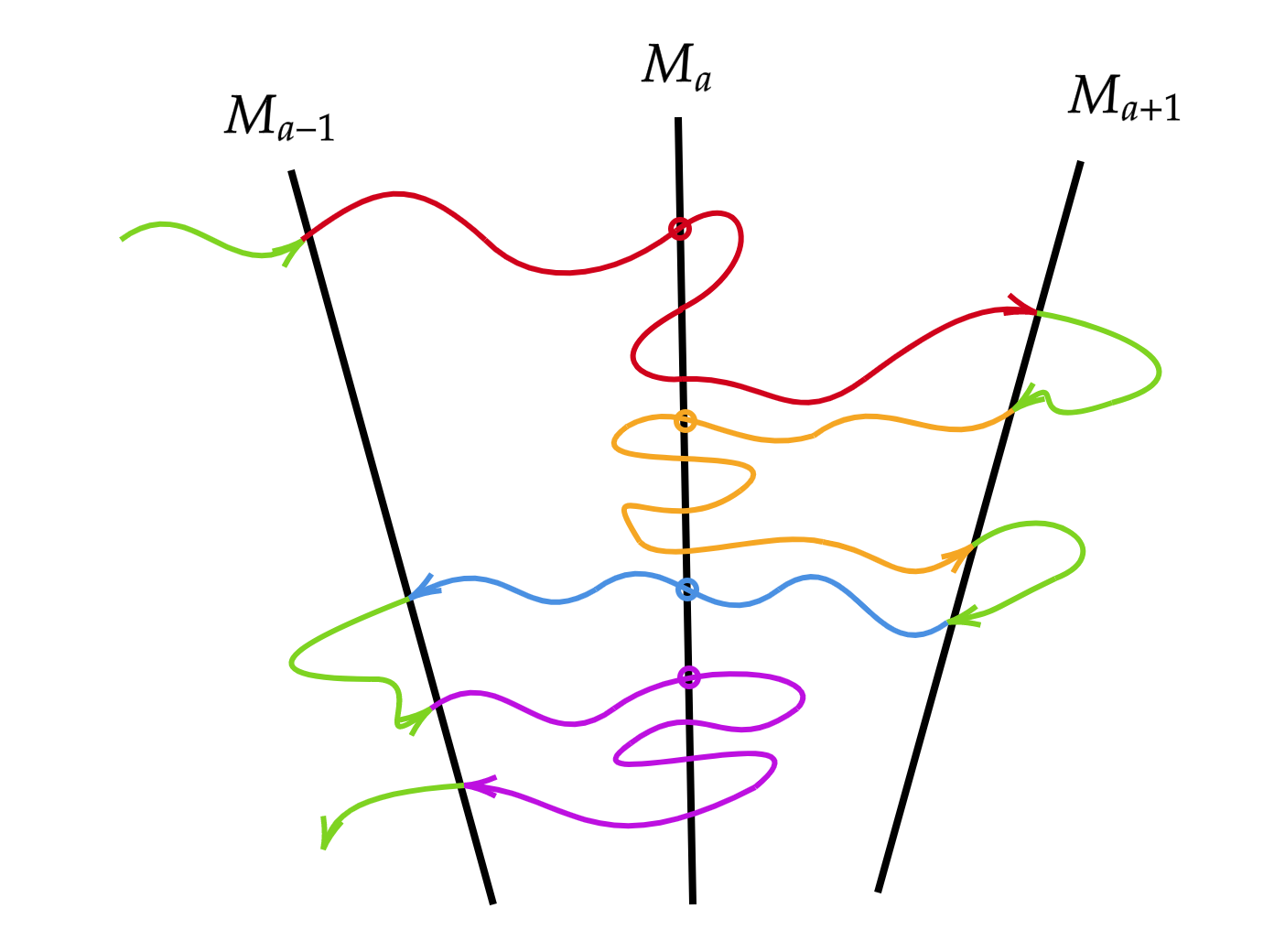}\\
\caption{Sketch of an equilibrium long trajectory passing through three consecutive milestones, $M_{a-1}$, $M_a$ and $M_{a+1}$. First hitting points on $M_a$ are marked in circles.}\label{fig_sketch 1}
\end{figure}

Assume after running an equilibrium long trajectory simulation we observe four transition events between milestones $M_{a-1}$ and $M_{a+1}$ passing through $M_a$ (see Fig. \ref{fig_sketch 1}). The transition probability is then estimated as $K_{M_a M_{a+1}}=2/4=0.5$ and $K_{M_a M_{a-1}}=2/4=0.5$ with each short segment of trajectories between $M_{a-1}$ and $M_{a+1}$ counted only once.
However, when we sample from BD on $M_a$ consisting of all crossing points, those short segments of trajectories consisting of more points within $[m,m+dM]$ will have a higher probability of being selected. 
The nonuniform (unnormalized) probability of the $j$th trajectory being selected from an initial BD sampling on $M_a$ is $p_j=\sum_ldt_{j,l}=\sum_l|v_{j,l}^\perp|^{-1}$, where the summation is over all crossing events of the $j$th trajectory on $M_a$. Here, $dt_{j,l}$ and $v_{j,l}^\perp$ are the local passage time and the velocity component normal to $M_a$ for the $j$th trajectory at the $l$th crossing event, respectively. 
Since each short segment of trajectories between $M_{a-1}$ and $M_{a+1}$ only contributes one first hitting point on $M_a$, the unbiased equilibrium FHPD derived from an initial BD sampling is estimated as 
\begin{equation}
f_{M_a}(x)=\frac{1}{Q}\sum_j \frac{1}{p_j}\delta(x-x_{FHP,j}),
\end{equation}
where the summation is over all short trajectories passing through $M_a$, $x_{FHP,j}$ is the first hitting point of the $j$th trajectory on $M_a$, and $Q=\sum_j \frac{1}{p_j}$ is the normalization factor.

Based on the this analysis, we devise the local passage time weighted Milestoning (LPT-M) algorithm. This algorithm constructs an approximate equilibrium FHPD, $f(x)$, from configurations sampled in the canonical ensemble as is done in CM but accounting for nonuniform trajectory weights. The algorithm is summarized as follows,

\begin{enumerate}[(1)]
\item Generate phase space configurations on each milestone in the canonical ensemble using restrained MD simulations, e.g., adding a harmonic restraint $\frac{1}{2}k(M(r^{3N})-m)^2$.

\item For each phase space configuration, run an unbiased trajectory forward in time until it hits a different milestone for the first time. Save the lifetime of the $j$th trajectory as $t_j^f$.

\item Reverse initial velocities and run an unbiased trajectory until it hits a different milestone for the first time. Save the lifetime from the initial configuration up to the last recrossing point of the $j$th trajectory as $t_j^b$. Optionally, save the configuration (and reversed velocity) at the last crossing point for the estimation $f(x)$.

\item During the $j$th trajectory running in both forward and backward direction, record the velocity component perpendicular to the initial milestone at each crossing point, denoted as $|v_{j,l}^\perp|$.

\item Perform normal Milestoning analysis, but now account for the nonuniform weight of each trajectory $w_j=(\sum_l|v_{j,l}^\perp|^{-1})^{-1}$. Here, the summation is over all crossing points in both forward and backward time integration of the $j$th trajectory. The transition probability and the mean dwelling time are calculated as $K_{\alpha\beta}=\frac{\sum_{j\in(\alpha\rightarrow\beta)} w_j}{\sum_j w_j}$ and $t_\alpha =\frac{\sum_j w_j(t_j^f+t_j^b)}{\sum_j w_j}$, respectively.
\end{enumerate} 

It is worth noting that the time reversibility is utilized for backward propagation in (3). 

The weighting factor $(\sum_l|v_l^\perp|^{-1})^{-1}$ was previously used in the context of transition path sampling from a single dividing surface by Hummer\cite{Humm04}. The single dividing surface is a separatrix located in the transition state in a two-state model. To generate an unbiased trajectory ensemble passing through the surface consistent with equilibrium long trajectories, each trajectory in the ensemble needs to carry a weight to be consist with an equilibrium density on the surface. The transition path ensemble is then defined by those trajectories connecting reactant and product state considering proper weight.
Milestoning network (Fig. \ref{fig_sketch 1}) can be regarded as a union of "transition state model", in which a dividing surface (e.g., $M_a$) separating left milestone ($M_{a-1}$) from right milestone ($M_{a+1}$). 
One key difference from Hummer's method is that not only transition paths connecting reactant and product (e.g., red and blue trajectories) contribute to the flux calculation, but also those returning back (e.g., orange and purple trajectories) play a role. 

\subsubsection{Bayesian Inference} \label{BI}
Now, let us consider a continuous equilibrium long trajectory where the first hitting points fall exactly on the milestone surface. This picture is conceptually different from LPT-M. 

We begin with BD sampling that is rigidly constrained to a milestone and construct an algorithm for approximating equilibrium FHPD using Bayesian inference,
\begin{equation}
P(x|FHP)=\frac{P_{eq}(x)P(FHP|x)}{P(FHP)},
\end{equation}
where $P_{eq}(x)\sim e^{-\beta H(x)}\delta(M(r^{3N})-m)$ is BD constrained to a milestone, $P(FHP|x)$ is the conditional probability that a given phase space configuration is a first hitting point, $P(FHP)$ acts as a normalization constant, and $P(x|FHP)$ is the equilibrium FHPD we are seeking.
To determine whether a given phase space configuration $x$ is a first hitting point, we can run the trajectory backward in time starting from $x$ and check if it recrosses the initial milestone.
For deterministic Hamiltonian dynamics $P(FHP|x)=\delta_{x,x_{FHP}}$ leading naturally to the DiM method\cite{DiM10}. 
For stochastic dynamics, such as Langevin dynamics, $P(FHP|x)$ is no longer a Kronecker delta function but a probability ratio. 

This lays the groundwork for the following Bayesian inference Milestoning (BI-M) algorithm,

\begin{enumerate}[(1)]
\item Generate configurations on each milestone in the canonical ensemble by constrained MD simulations.

\item For each configuration, draw initial velocities from the Maxwell distribution. 

\item For each configuration, reverse the initial velocity and run an unbiased trajectory until it hits a different milestone for the first time or recrosses the initial milestone, whichever occurs first. Repeat this procedure for $n$ times for each configuration, and count the number of times $n_j$ when the initial phase space configuration is a first hitting point. Estimate $P(FHP|x_j)=n_j/n$.

\item For those initial phase space configurations with non-zero $P(FHP|x_j)$, run an unbiased trajectory forward in time until it hits a different milestone for the first time. Record the trajectory lifetime $t_j$.

\item Perform normal Milestoning analysis but now considering the nonuniform weight of each trajectory $w_j= z(r^{3N})^{-\frac{1}{2}}P(FHP|x_j)$. The transition probability and the mean dwelling time are calculated as $K_{\alpha\beta}=\frac{\sum_{j\in(\alpha\rightarrow\beta)} w_j}{\sum_j w_j}$ and $t_\alpha =\frac{\sum_j w_jt_j}{\sum_j w_j}$, respectively. 

\end{enumerate}  

In (1), we assume that the initial configurations sampled fall exactly on milestones for estimating $P(FHP|x)$. This can be achieved by imposing a rigid constraint $\sigma(r^{3N})=M(r^{3N})-m=0$ using Lagrange multiplier or SHAKE algorithm\cite{SHAKE}. When a rigid constraint is imposed\cite{ConsMD98}, the configurations sampled in (1) obey the distribution $f(r^{3N})\sim z(r^{3N})^{\frac{1}{2}}e^{-\beta U(r^{3N})}\delta(M(r^{3N})-m)$ with the metric factor $z(r^{3N})=\sum_{i=1}^{N}\frac{1}{m_i}(\frac{\partial\sigma}{\partial\mathbf{r}_i})^2$. Consequently, we include an additional factor $z(r^{3N})^{-\frac{1}{2}}$ in the weight to balance it.

The choice of the number of backward trajectories, $n$, to run in (3) determines the resolution of the conditional probability $P(FHP|x_j)$. In the current study, we choose $n=10$, indicating the smallest discernible change in $P(FHP|x_j)$ is $0.1$. During our numerical tests, we observed no change in MFPT by increasing $n$ to 20.  

\subsubsection{Locally Iterative Correction} \label{LiC}

Due to the difference between the equilibrium FHPD and the non-equilibrium stationary FHPD of the partial path ensemble $A\rightarrow B$, MFPT calculated using LPT-M and BI-M methods are not exact. 
We here introduce a ratio function $R_{A\rightarrow B}$ as an indicator to measure the contribution of the path ensemble $A\rightarrow B$ to the equilibrium FHPD on a milestone. When $R_{A\rightarrow B}$ falls below a certain threshold (e.g., $80\%$), it suggests a potential significant discrepancy between the two FHPDs on a milestone. In such cases, an iterative correction (following Eq. \eqref{FHPD iter}) can be specifically applied to those problematic milestones (i.e., locally iterative correction). 
This approach allows for a substantial reduction in computational cost in each iteration compared to the ExM method.

The iterative correction to problematic milestones follows the ExM algorithm but generalizes it to nonuniform trajectory weights. In step (iv) of the ExM method, where FHPD is prepared for the next iteration, we resample first hitting points using their assigned weights from LPT-M or BI-M methods, rather than using a uniform sampling as in ExM. 

Let us illustrate the procedure of computing $R_{A\rightarrow B}$ for each milestone.
Given a full transition probability matrix $\mathbf{K}$ obtained from either LPT-M or BI-M, we reduce it to a $3\times 3$ transition probability matrix $\mathbf{K}^R$ with row 1 corresponding to the reactant milestone $\xi$, row 2 corresponding to an intermediate target milestone $\alpha$ for which we want to calculate $R_{A\rightarrow B}$, and row 3 corresponding to the product milestone $\gamma$,
\begin{equation}
\mathbf{K}^R=
\begin{bmatrix}
0 & K^R_{12} & K^R_{13} \\
K^R_{21} & 0 & K^R_{23} \\
K^R_{31} & K^R_{32} & 0 
\end{bmatrix}.  
\end{equation}

Here, the matrix element $K^R_{21}$ represents the commitment probability (or committor) that a trajectory initiated on the target milestone $\alpha$ will reach the reactant milestone $\xi$ first before the product milestone $\gamma$. The commitment probability corresponds to the splitting probability introduced by Onsager for ion-pair recombination\cite{SplitProb}. 
This reduction of the transition probability matrix preserves the effective transition probabilities between milestones and mains the stationary flux through milestones.
After this reduction, the reactant (and product) milestones and the target milestone are directly connected, which makes the calculation of $R_{A\rightarrow B}$ for the target milestone more straightforward.

It is noteworthy that the committor function on a milestone $C_\alpha(x_\alpha)$ is in general not a constant, unless the milestone is an isocommittor surface. The committor functions are computationally expensive and approximations have to be adopted in practice. Here, we assume that the committor value on milestones is constant, $C_\alpha(x_\alpha)=C_\alpha$. This assumption is reasonable for two reasons: (i) this is approximately true when the milestone surface is of small size, which is usually the case, such that $C_\alpha(x_\alpha)$ does not change significantly on a milestone; (ii) Our final goal is a cheap and rough estimation of $R_{A\rightarrow B}$ rather than a highly accurate result. This constant value assumption turns out to be good enough for our purpose. 

With the constant committor value assumption in mind, calculating committors in Milestoning becomes straightforward\cite{CommMile}. To do this, we select two out of the three milestones within $\mathbf{K}^R$, say $\alpha$ and $\gamma$, as the two end states. Using the first-step analysis, it can be readily verified that the committor $C_\beta$ of any milestone $\beta$, $\beta\in\{M_1,M_2,\cdots,M_n\}$, to first reach $\alpha$ before reaching $\gamma$ satisfies the following equation,
\begin{equation}
(\mathbf{I}-\tilde{\mathbf{K}})\mathbf{C}=\mathbf{e}_\alpha,\label{committor eq}
\end{equation}
where $\mathbf{I}$ is an $n\times n$ identity matrix, $\tilde{\mathbf{K}}$ is the full transition probability matrix with the exception of setting two rows corresponding to milestones $\alpha$ and $\gamma$ to zero, i.e., $\tilde{K}_{\alpha\beta}=0$ and $\tilde{K}_{\gamma\beta}=0$ for all $\beta\in\{M_1,M_2,\cdots,M_n\}$, $\mathbf{C}$ is a column vector of committors, and $\mathbf{e}_\alpha$ is a column vector with all elements being zero except for the element corresponding to $\alpha$, which is set to one. After solving Eq. \eqref{committor eq} for $\mathbf{C}$, we set $K^R_{12}=C_\xi$.
This process is repeated until all the matrix elements in $\mathbf{K}^R$ are obtained.

Finally, we solve the eigenvalue equation $\mathbf{q}^T=\mathbf{q}^T\mathbf{K}^R$ to obtain the equilibrium flux. The flux through the target milestone combines contributions from both the reactant and product milestones, $q_2=K^R_{12}q_1 + K^R_{32}q_3$. Consequently, the ratio of the contribution of the partial path ensemble $A\rightarrow B$ to equilibrium FHPD on the target milestone $\alpha$ can be calculated as $R_{A\rightarrow B}(\alpha)=q_1K^R_{12}/q_2$. 

\subsection{Simulation Details} \label{SimuDeta}
\emph{Mueller's Potential.} The Mueller's potential is a 2D model system that has been used for benchmark test of kinetics\cite{Tilting09,MarkovM09,MileW20}. Voronoi tessellation is employed to partition the configuration space into small cells, as illustrated in Fig. \ref{fig_muller}.
These cells, denoted as $B_i$, are defined by
\begin{equation}
B_i = \{(\mathbf{r},\mathbf{v})\in\mathbb{R}^2\times\mathbb{R}^2: \|\mathbf{r}-\mathbf{r}_i\|_2<\|\mathbf{r}-\mathbf{r}_j\|_2\quad \mathrm{for}\quad \mathrm{all}\quad j\neq i\}.
\end{equation}
The Voronoi centers $\{\mathbf{r}_i\}$, also called anchors, are placed along the minimum energy pathway (MEP) optimized by the zero-temperature string method\cite{String02}. Underdamped Langevin dynamics is employed for simulations,
\begin{align}
\dot{\mathbf{r}} &= \mathbf{v}, \\
m\dot{\mathbf{v}} &= -\frac{\partial U}{\partial \mathbf{r}} - \Gamma \mathbf{v} + \mathbf{\eta}.
\end{align}
The Euler-Maruyama algorithm is utilized with the integration time step $\Delta t=10^{-4}$, temperature $k_BT=10$, friction coefficient $\Gamma=10$ (or $100$), and mass $m=1$. The white noise $\mathbf{\eta}(t)$ is of mean zero and covariance $\langle\eta_i(t)\eta_j(t')\rangle=2\Gamma k_BT\delta_{ij}\delta(t-t')$. The energy barrier along the MEP is approximately $10 k_BT$.

In LPT-M and CM methods, harmonic restraints are used to initially sample on each milestone, whereas in BI-M, a rigid constraint is applied for initial sampling. Implementing the rigid constraint within Voronoi tessellation involves projecting initial velocities and forces at each time step onto the milestone hyperplane\cite{ConstLD11}. 
In BI-M, from each initially sampled phase space configuration, ten backward trajectories are run to estimate $P(FHP|\mathbf{r},\mathbf{v})$ in BI-M.
The metric factor $z(\mathbf{r})$ associated with the rigid constraint $\sigma(\mathbf{r})=(\mathbf{r}-\mathbf{r}_i)^2-(\mathbf{r}-\mathbf{r}_j)^2=0$ is a constant. Therefore, the weighting factor in BI-M is directly $w=P(FHP|\mathbf{r},\mathbf{v})$. Occasionally, trajectories generated by forward and backward time integration in LPT-M have no crossings with the initial milestone. Such trajectories are disregarded in the Milestoning analysis. A total of $1000$ effective trajectories are run from each milestone, with ten independent simulations conducted for each method (LPT-M, BI-M and CM). The average and standard deviation of MFPT are reported. 

The locally iterative LPT-M (LiLPT-M) method is performed as described in Sec. \ref{LiC}. All milestones with $R_{A\rightarrow B}$ below $80\%$ participate in the iteration. Five independent simulations are conducted for error estimation. The transition probability and mean dwelling time of those milestones that do not need correction are averaged first before iteration starts. 

\emph{Deca-alanine Unfolding in Vacuum.} The system is modeled as ACE-(Ala)$_{10}$-NME. The NAMD 2.14 program\cite{NAMD} and CHARMM36 force fields\cite{CHARMM36} are used for MD simulations. The integration time step is $1$ fs. All atoms are included in nonbond interactions, i.e., no cutoff distance is set. The system runs in NVT ensemble using a Langevin thermostat at 600K with a friction constant $1$ ps$^{-1}$. The end-to-end distance between two carbon atoms $d_{CC}$ is used as the reaction coordinate to characterize the unfolding process. The reactant (folded) and product (fully extended) state are defined as $d_{CC}=13$ \AA\ and $d_{CC}=34$ \AA, respectively. A total of $22$ milestones uniformly separated by $1$ \AA\ are placed along $d_{CC}$ in $[13$ \AA $,34$\AA$]$.

A total of $400$ configurations are sampled in the canonical ensemble at the reactant by restrained MD simulations. Long trajectory ensemble are then run uninterrupted from the reactant to the product state. The time average is used as the MFPT reference. 

CM and LPT-M calculations employ restrained MD simulations with a force constant $12$ kcal/mol/\AA$^2$ for initial sampling on each milestone. The velocity component normal to a milestone is calculated using finite difference as $|d_{CC}(t+\Delta t)-d_{CC}(t)|/\Delta t$ with $t$ and $t+\Delta t$ being the time moment right before and after the crossing, respectively. In BI-M calculations, constrained MD simulations are used for initial sampling on each milestone by fixing the positions of the two carbon atoms at the ends. Configurations are stored every $2$ ps.
In BI-M, ten backward trajectories are run from each initially sampled phase space configuration to estimate $P(FHP|\mathbf{r},\mathbf{v})$. The metric factor $z(\mathbf{r})$ associated with the rigid constraint $\sigma(\mathbf{r}_{C_1},\mathbf{r}_{C_2})=\|\mathbf{r}_{C_1}-\mathbf{r}_{C_2}\|_2-d_{CC}^0=0$ is a constant. A total of $400$ effective trajectories are run from each milestone for Milestoning calculations. Five independent simulations are conducted for each method (LPT-M, BI-M and CM). The average and standard deviation of MFPT are reported.

\section{Results and Discussions}\label{Results}
In this section, we assess the performance of LPT-M, BI-M and LiLPT-M algorithms using two model systems: Mueller's potential and deca-alanine unfolding in vacuum.

The performance is evaluated from two perspectives: (i) the accuracy of predicted MFPT; (ii) the sensitivity of predicted MFPT with respect to the number of milestones. 
We gauge the accuracy of the methods by examining their predicted MFPT. The exact MFPT can be attained when the non-equilibrium stationary FHPD of the partial path ensemble $A\rightarrow B$ is known. In such cases, the configuration space can be partitioned in a finely detailed manner, aiming for maximum efficiency without compromising accuracy. As a result, the sensitivity of methods to the number of milestones employed shed light on how well the non-equilibrium stationary FHPD is approximated.

\subsection{Mueller's Potential}\label{Muller's potential} 
\begin{figure}[h]{}
\centering
\begin{tabular}{cc}
\includegraphics[height=7cm]{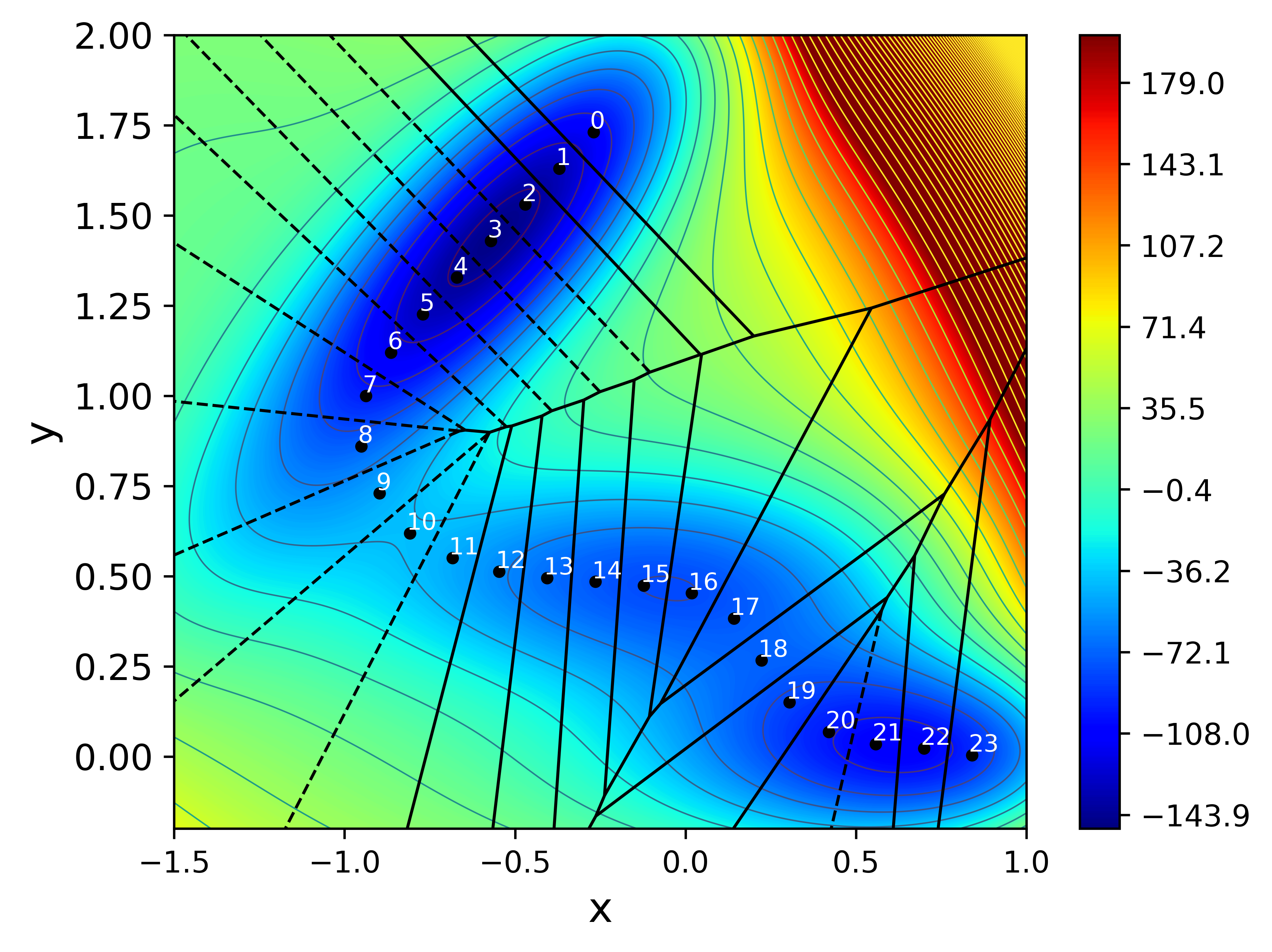}\\
(a)\\
\includegraphics[height=7cm]{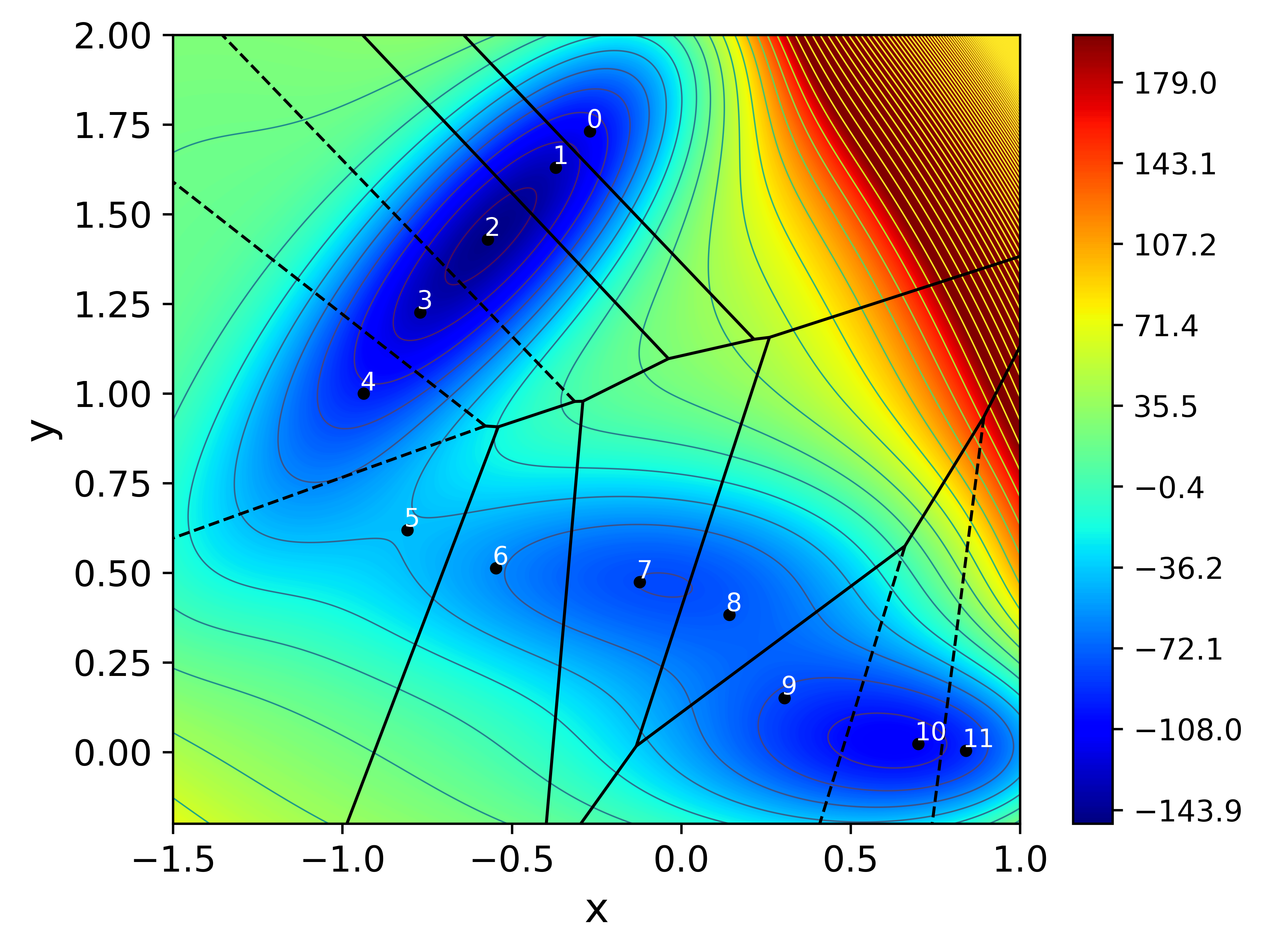}\\
(b)
\end{tabular}
\caption{Voronoi tessellation on the Mueller's potential with (a) 24 or (b) 12 anchors placed along the minimum energy pathway.}\label{fig_muller}
\end{figure}

For the low friction case with $\Gamma=10$, where the inertia effect is significant, we place 24 anchors along the MEP (Fig. \ref{fig_muller} (a) and Table S1). Each milestone (cell interface) is indexed by two anchors defining it in Voronoi tessellation. The milestones $(0,1)$ and $(22,23)$ are defined as the reactant and the product state, respectively. 

\begin{figure}[h]
\centering
\begin{tabular}{cc}
\includegraphics[height=7cm]{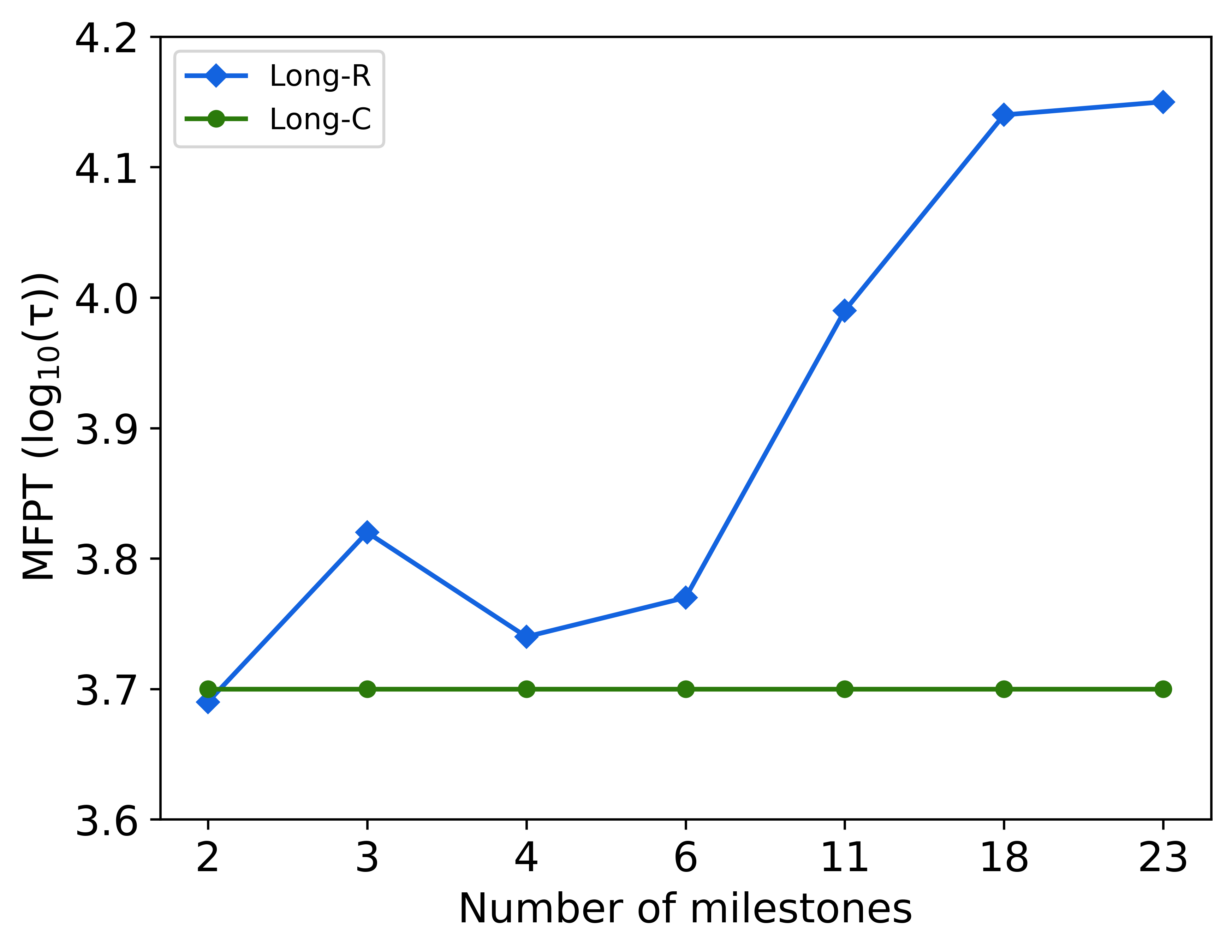}\\
(a)\\
\includegraphics[height=7cm]{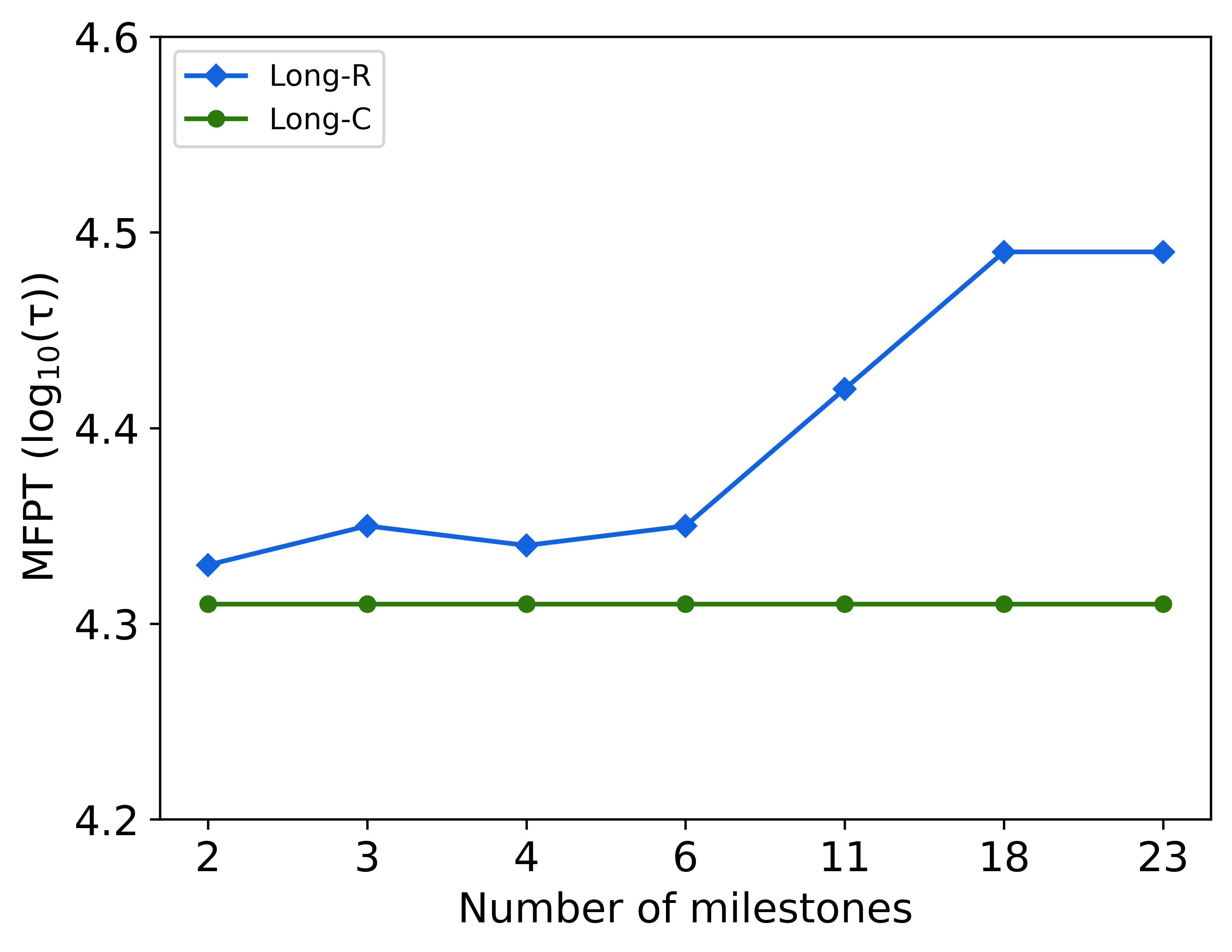}\\
(b)
\end{tabular}
\caption{MFPT calculations from a long trajectory simulation with (a) $\Gamma=10$ or (b) $\Gamma=100$ using reflecting boundary condition (Long-R) or cyclic boundary condition (Long-C) at product on Mueller's potential. Milestoning analysis is performed with increasing number of intermediate milestones.}\label{fig_mfpt_long}
\end{figure}

We first compare Milestoning analysis combined with a long trajectory simulation in two different boundary conditions at the product. One uses reflecting boundary condition (time length $1\times 10^7$), while the other uses cyclic boundary condition ($1000$ unidirectional transitions from reactant to product with a total time length of about $5\times 10^6$). 
We gradually increase the number of milestones from 2 to 23 along the MEP during Milestoning analysis (Fig. \ref{fig_mfpt_long} (a) and Table S3). 

When only minimally two milestones (reactant and product) are retained, omitting all intermediate milestones, both boundary conditions yield exact MFPT. This is because the transition probability matrices in both boundary conditions now have the same form,     
\begin{equation}
\mathbf{K}=
\begin{bmatrix}
0 & 1  \\
1 & 0  \\ 
\end{bmatrix}.  
\end{equation}
Upon substituting the above $\mathbf{K}$ matrix into Eq. \eqref{flux cyc} and solving Eq. \eqref{MFPT cyc} for MFPT, it becomes apparent that the MFPT corresponds to the lifetime of the reactant milestone. Given that the $A\rightarrow B$ transition is a rare event, the lifetime of the reactant milestone is insensitive to the initial distribution of $A$ in the two-state scenario. Therefore, the lifetime of $A$ is the same in both boundary conditions.

As intermediate milestones are gradually introduced, MFPT prediction with the reflecting boundary condition starts to deviate from the reference value, particularly with 23 milestones. In contrast, the MFPT prediction with cyclic boundary condition remains accurate and stable. 
This confirms our statement that the true initial distribution of the short trajectory ensemble required for exact MFPT calculation is the non-equilibrium stationary FHPD of the partial path ensemble $A\rightarrow B$. 

\begin{figure}[h]
\centering
\begin{tabular}{cc}
\includegraphics[height=7cm]{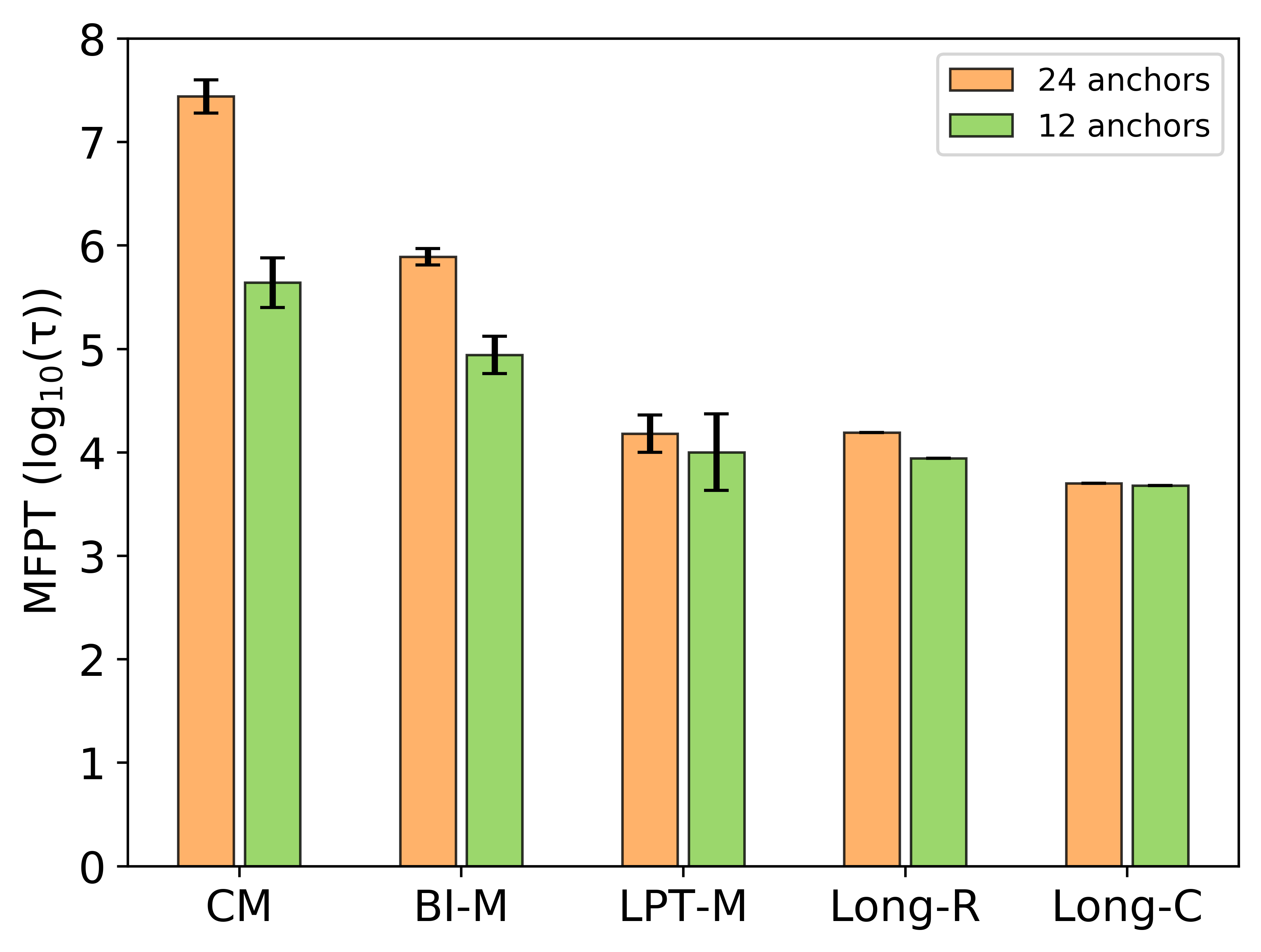}\\
(a)\\
\includegraphics[height=7cm]{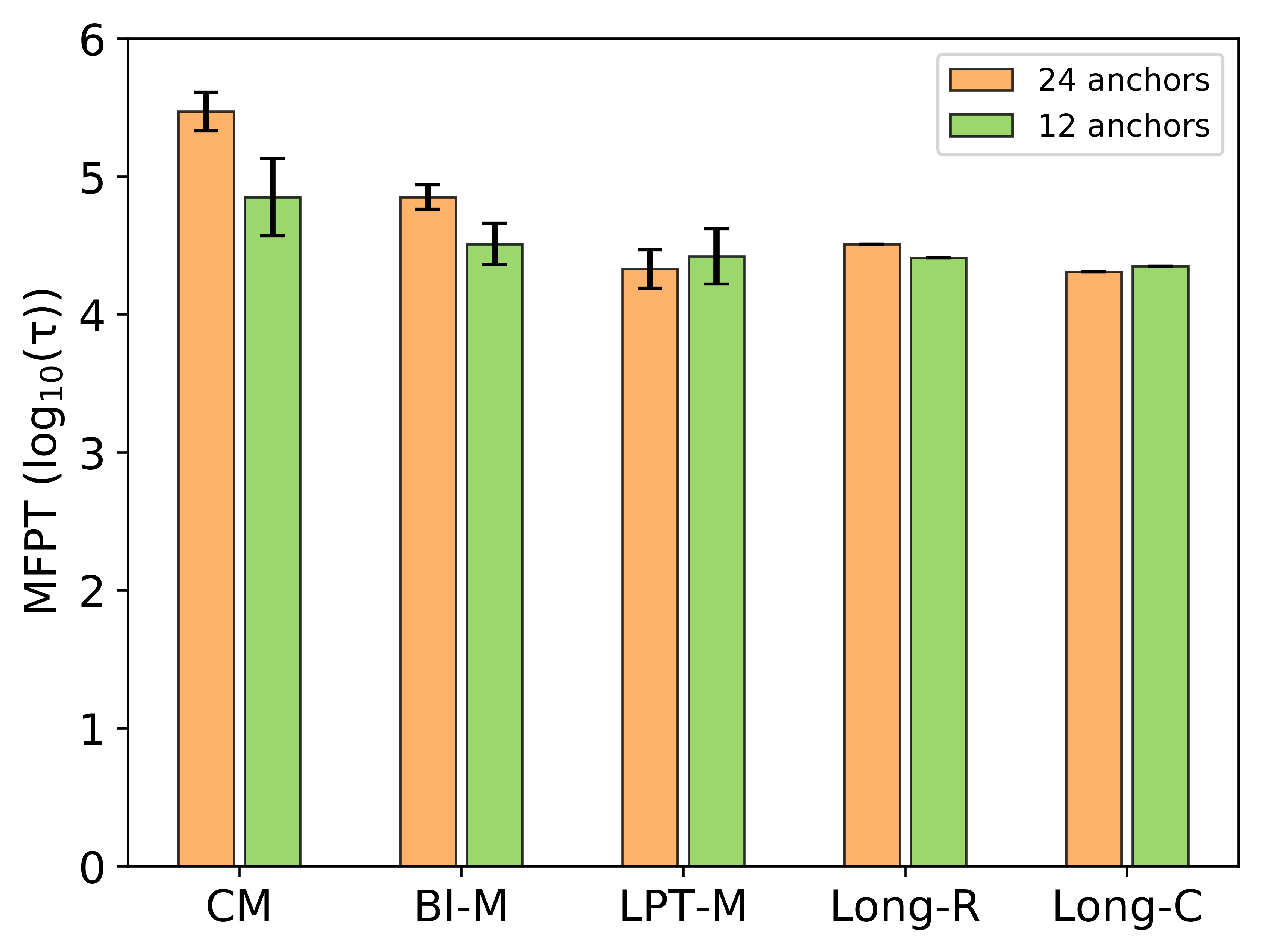}\\
(b)
\end{tabular}
\caption{MFPT calculations obtained from classical Milestoning (CM), BI-M, LPT-M, and a long trajectory simulation with reflecting boundary condition (Long-R) or cyclic boundary condition (Long-C). 12 or 24 anchors are placed along the minimum energy pathway on Mueller's potential. (a) $\Gamma=10$, (b) $\Gamma=100$.}\label{fig_mfpt}
\end{figure}

We proceed to compare LPT-M and BI-M with the CM method in both dense (24 anchors) and sparse (12 anchors) partition of the configuration space (Fig. \ref{fig_mfpt} (a)). To accentuate the contrast, the MFPT values for each method are listed in Table S4. 
The reference for these comparisons is the the long trajectory simulation result with cyclic boundary condition. 
The results underscore the significant improvement of both LPT-M and BI-M over CM. It is noteworthy that the LPT-M method approaches the accuracy limit of the equilibrium long trajectory simulation with reflecting boundary condition. In particular, LPT-M exhibits reduced sensitivity to an increasing number of intermediate milestones.

The computational costs are evaluated in terms of the number of force builds per milestone ($N_\mathrm{FB}$), as force evaluation is the most time-consuming part in MD simulations. 
The count of $N_{\mathrm{FB}}$ includes both the initial restrained sampling and subsequent free evolution of short trajectory ensemble. 
As summarized in Table \ref{Nfb muller}, LPT-M (BI-M) is about 1.5 (5.0) times as expensive as CM in both dense and sparse partition of the configuration space. However, it should be noted that the forward and backward evolution in LPT-M can run independently, resulting in the same wall-clock time as CM. The number of force builds of LPT-M and BI-M is reduced by over three orders of magnitude compared to direct long trajectory ensemble simulations. 

The errors in transition probability and mean dwelling time for milestones along the MEP are compared in Fig. S1 and S2, respectively. The errors predominantly manifest in the second half of MEP. It is noteworthy that milestones close to the reactant exert a more significant influence on MFPT calculations due to their significantly larger flux compared to those near the product (cf. Eq. \eqref{MFPT cyc}). By comparison with the $R_{A\rightarrow B}$ profile (Fig. S3), the conclusion can be drawn that, under the current set-up, when $R_{A\rightarrow B}$ falls below $80\%$, the discrepancy between equilibrium FHPD and non-equilibrium stationary FHPD of the partial path ensemble $A\rightarrow B$ becomes significant.  

\begin{figure}[h]
\centering
\begin{tabular}{cc}
\includegraphics[height=7cm]{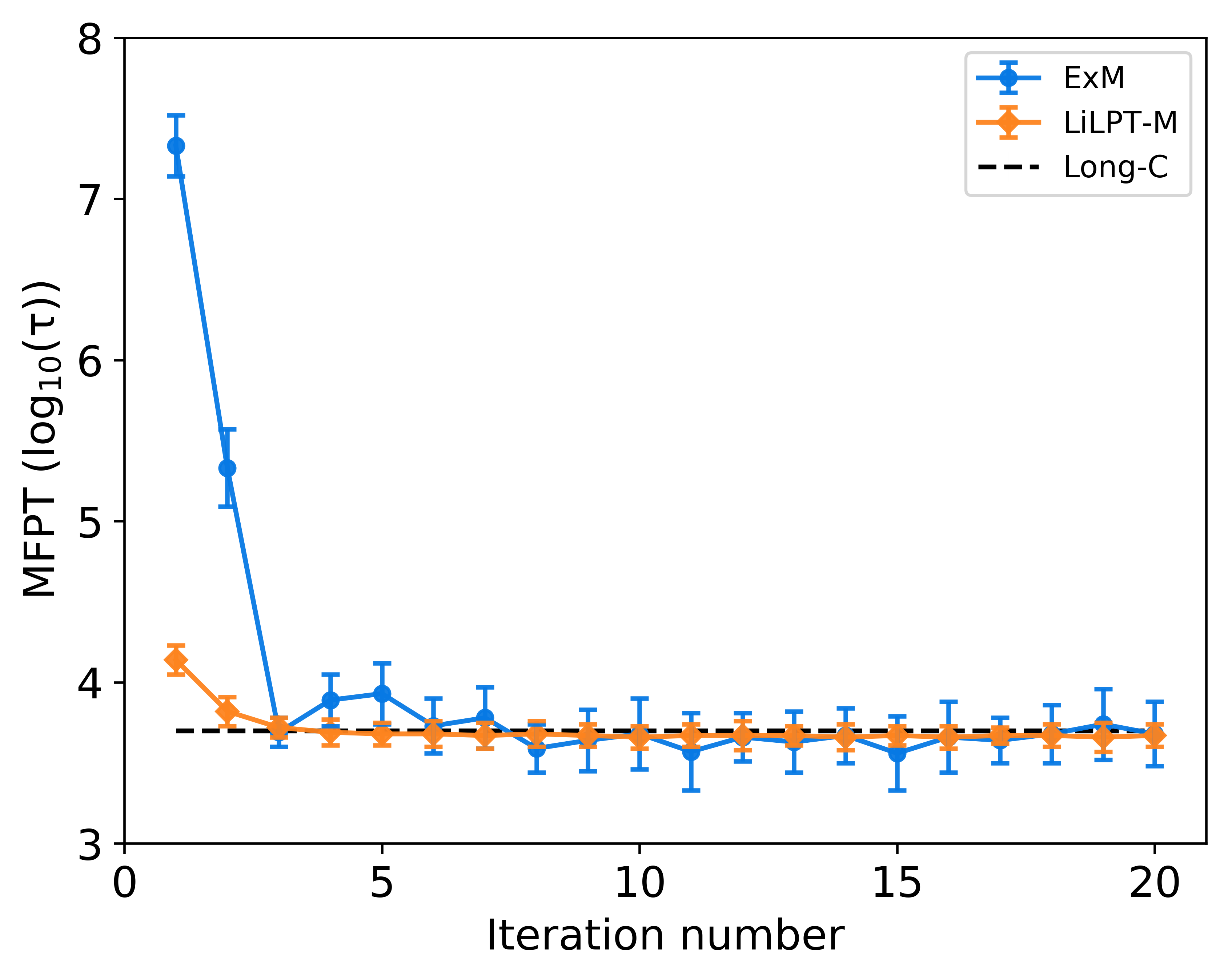}\\
(a)\\
\includegraphics[height=7cm]{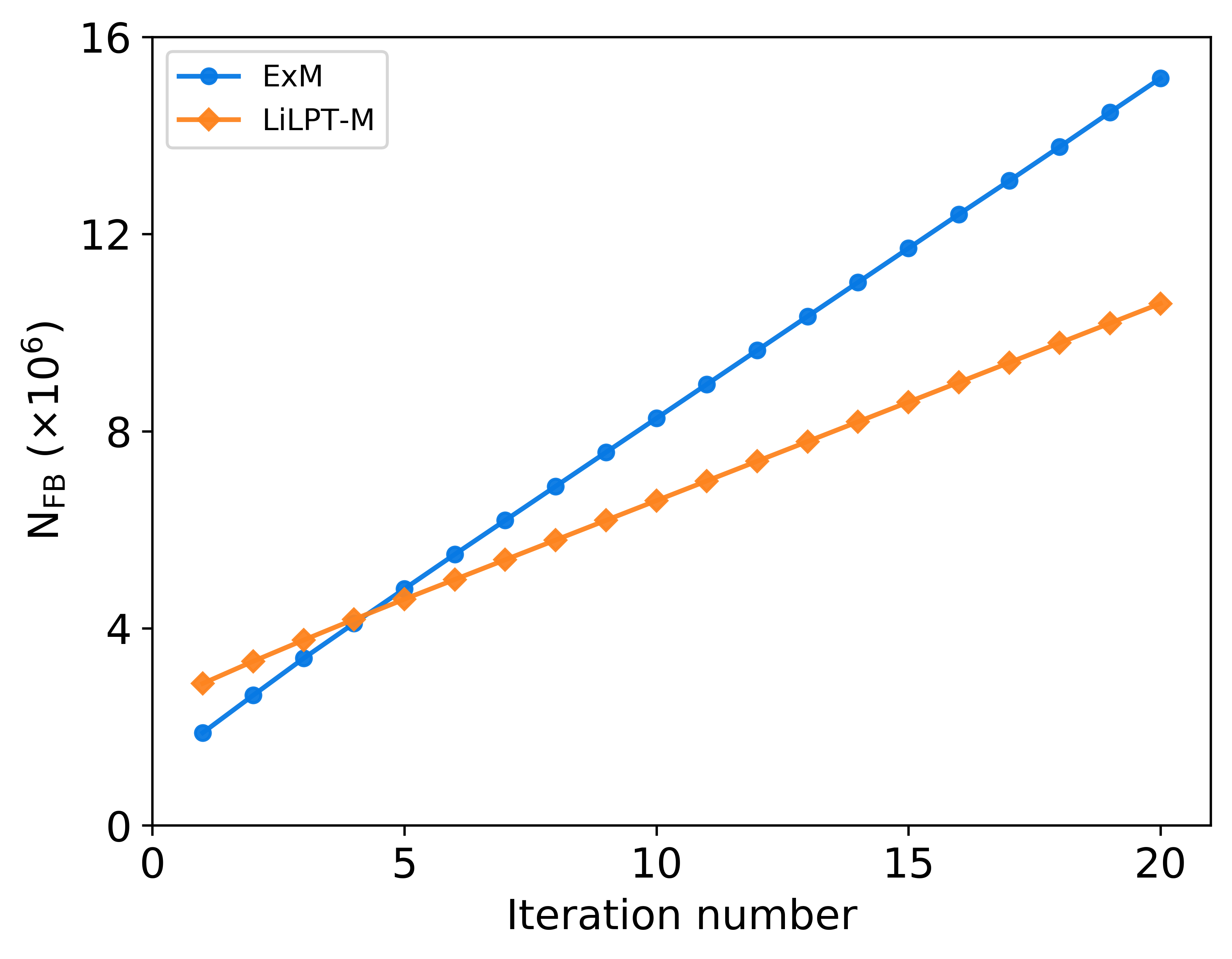}\\
(b)
\end{tabular}
\caption{The Mueller's potential is partitioned using 24 anchors along the minimum energy pathway. Dynamics is evolved with $\Gamma=10$. (a) MFPT from locally iterative LPT-M (LiLPT-M) and exact Milestoning (ExM) method are compared. MFPT from a long trajectory simulation with cyclic boundary condition (Long-C) at product is used as reference. (b) Computational costs in terms of the number of force builds per milestone $N_\mathrm{FB}$ for LiLPT-M and ExM.}\label{fig_iter}
\end{figure}

LiLPT-M and ExM are compared in Fig. \ref{fig_iter} and Table S5. LiLPT-M converges at the third iteration (stably within $10\%$ of the reference MFPT value), while ExM only roughly converges at the 9th iteration. LiLPT-M converges more rapidly and exhibits better stability during the iteration process. Since only a subset of milestones participate in the iteration, its computational cost increases at a lower rate. At convergence, ExM is about twice as expensive as LiLPT-M.

In 2007, West, Shalloway, and Elber argued that CM gives adequate MFPT and free energy when the velocity correlation functions decrease to zero between milestones\cite{Elber07}. They illustrate this result for solvated alanine dipeptide. In the limit of $\Gamma=10$ for the Mueller's potential, the velocity is not relaxing to zero between milestones. Therefore CM is not adequate for this system.

We next consider $\Gamma=100$, which represents a moderate friction case. The randomness of trajectories has increased compared to that of $\Gamma=10$. 
The MFPT from an equilibrium long trajectory simulation with reflecting boundary condition (time length $5\times 10^7$) and cyclic boundary condition ($1000$ unidirectional transitions from reactant to product with a total time length of about $2\times 10^7$) is shown in Fig. \ref{fig_mfpt_long} (b). The MFPT calculation with cyclic boundary condition continues to exhibit robustness with respect to the increasing number of milestones. 
However, the MFPT calculation with reflecting boundary condition is now less sensitive to the increasing number of milestones, as the velocity decorrelation is faster in $\Gamma=100$ compared to $\Gamma=10$. The deviation of MFPT with 23 milestones is about 1.5 times as large as the reference.

The MFPT accuracy of CM, BI-M and LPT-M at $\Gamma=100$ is also compared (Fig. \ref{fig_mfpt} (b) and Table S4). At no surprise, all three methods are now closer to the MFPT reference, and LPT-M remains the most accurate. In addition, all the three methods are less sensitive to the increasing number of milestones, due to the faster velocity decorrelation. In Table \ref{Nfb muller}, their computational costs are evaluated. BI-M is about 3.3 times as expensive as CM, lower than the ratio at $\Gamma=10$. LPT-M is about 1.7 times as expensive as CM, with the ratio roughly the same as that in $\Gamma=10$. The number of force builds in LPT-M and BI-M is reduced by over three orders of magnitude compared to direct long trajectory ensemble simulations.

\begin{table}
\centering
\caption{Computational costs in terms of the number of force builds per milestone ($N_\mathrm{FB}$) obtained from CM, BI-M, LPT-M and a long trajectory simulation with cyclic boundary condition (Long-C) using 12 or 24 anchors along the minimum energy pathway on Mueller's potential.}\label{Nfb muller}
\begin{threeparttable}
\begin{tabular}{ccccc}\toprule
 $N_{\mathrm{FB}}$ ($\times 10^6$)  & \multicolumn{2}{c}{$\Gamma=10$} & \multicolumn{2}{c}{$\Gamma=100$}  \\
\hline
        &  $12$ anchors & $24$ anchors   & $12$ anchors & $24$ anchors \\
CM      &     3.16      & 1.88           & 6.79         & 2.67   \\
BI-M    &     17.04     & 8.18           & 22.62        & 8.45   \\
LPT-M   &     5.56      & 2.88           & 12.66        & 4.44  \\
Long-C  &     4363      & 2174           & 20909        & 8696 \\
\bottomrule
\end{tabular}
\end{threeparttable}
\end{table}

\subsection{Deca-alanine Unfolding in Vacuum}\label{Deca-alanine unfolding} 

\begin{figure}[h]
\centering
\begin{tabular}{cc}
\includegraphics[height=7cm]{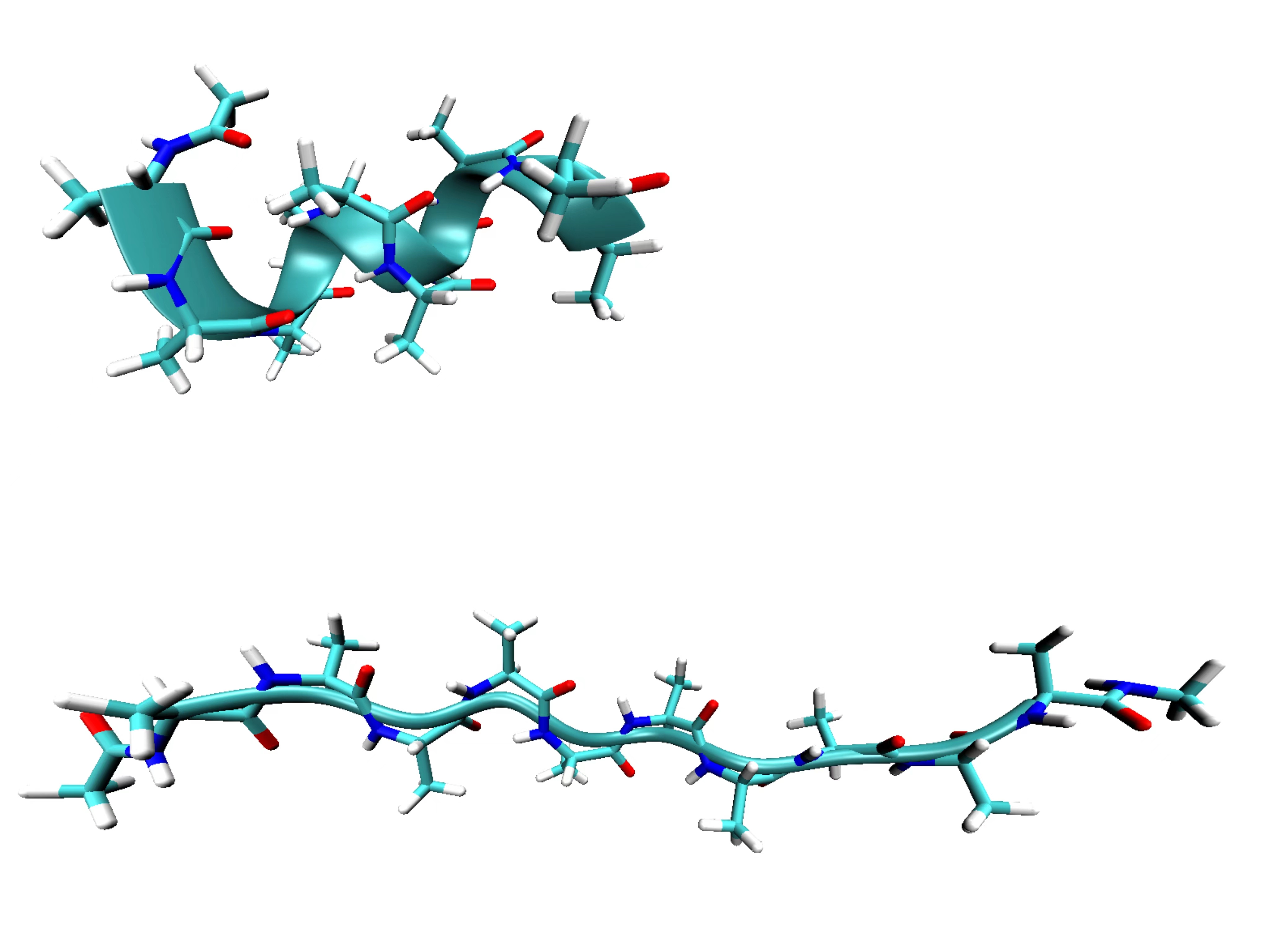}\\
(a)\\
\includegraphics[height=7cm]{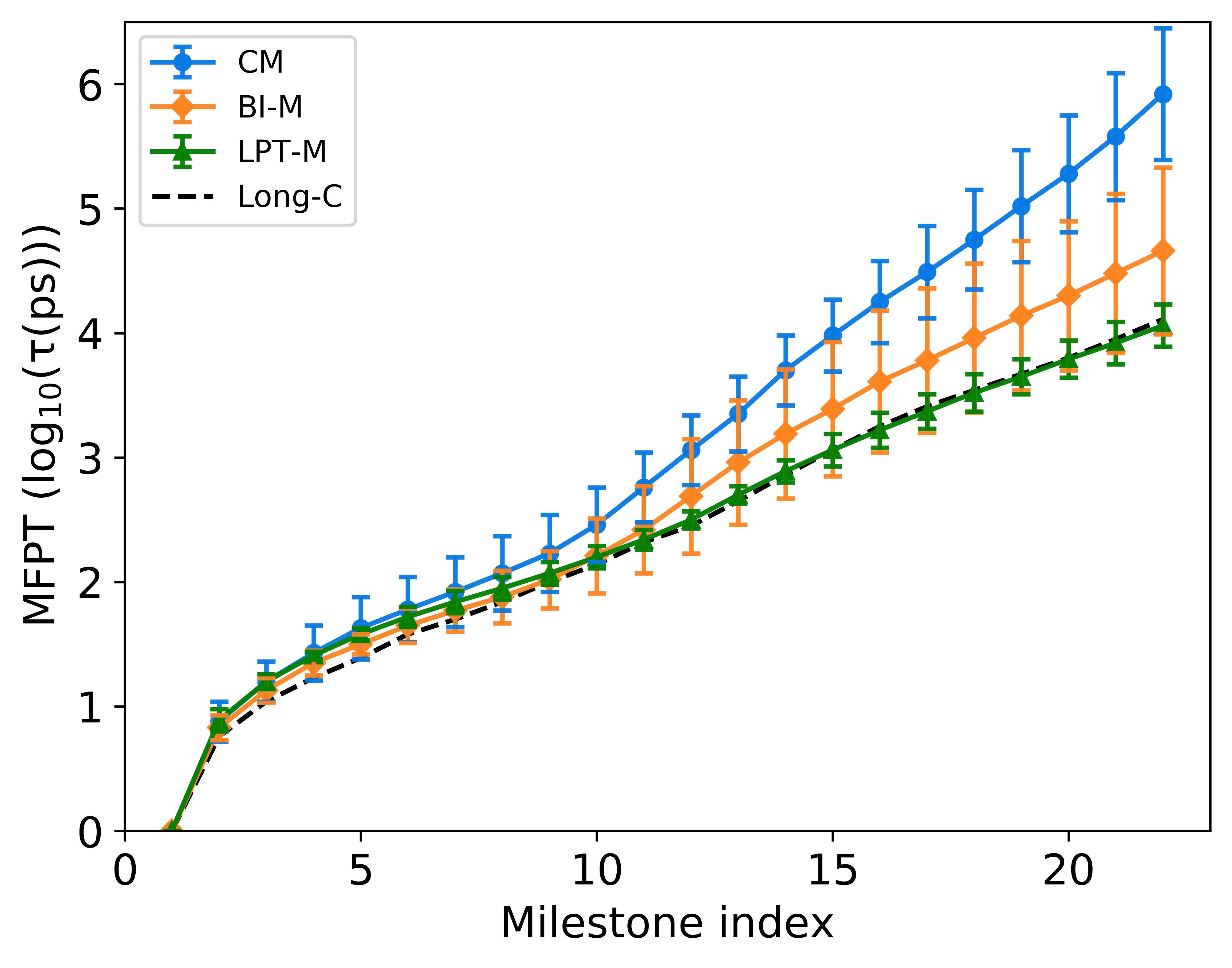}\\
(b)
\end{tabular}
\caption{(a) The folded and fully extended structure of deca-alanine. (b) MFPT of CM, LPT-M and BI-M from the folded state (milestone $1$) to each intermediate milestone until the fully extended one (milestone $22$) along the deca-alanine unfolding process in vacuum. MFPT from a long trajectory simulation with cyclic boundary condition (Long-C) serves as reference.}\label{fig_ala}
\end{figure}

MFPT results of CM, BI-M and LPT-M, originating from the folded state (reactant, Fig. \ref{fig_ala} (a)) and traversing each intermediate state along the unfolding pathway until reaching the fully extended state (product, Fig. \ref{fig_ala} (a)), are shown in Fig. \ref{fig_ala} (b). Both BI-M and LPT-M improve over CM, with LPT-M notably fitting remarkably well to the reference derived from the long trajectory ensemble with low variance. Table. \ref{Nfb unfolding} summarizes their computational costs in terms of the $N_{\mathrm{FB}}$. LPT-M is about $1.4$ times as expensive as CM, while BI-M is about $3.3$ times as expensive as CM. Both LPT-M and BI-M reduce the number of force builds by over two orders of magnitude compared to direct long trajectory ensemble simulations.

The errors of transition probabilities and mean dwelling time for milestones along the unfolding pathway are shown in Fig. S4 and S5. The result shows that the transition probability errors of LPT-M and BI-M are small on most milestones except the second to last. The $R_{A\rightarrow B}$ profile remains close to $100\%$ in the first half and falls quickly as approaching the product state (Fig. S6). In particular, $R_{A\rightarrow B}$ of the second to last milestone falls below $40\%$, which clearly raises a warning about potential discrepancy between the two FHPDs.

Throughout our investigation of both the Mueller's potential and deca-alanine unfolding, we consistently observe higher accuracy with LPT-M compared to BI-M. These two methods represent different conceptualizations of the initial FHPD. LPT-M assumes that all crossing points, including first hitting points, fall within a small interval near each milestone. Consequently, it initiates with harmonically restrained sampling. In contrast, BI-M assumes that all crossing points fall exactly on milestones, and therefore it initiates with rigidly constrained sampling. 
Notably, the assumption that crossing points can fall exactly on milestones clashes with the underlying logic of LPT-M.
In fact, the thickness of the interval $[m,m+dM]$ in LPT-M can be made arbitrarily small but never reduced to zero. In the latter case, the local passage time also approaches zero, leading to a divergent weighting factor. 

Considering transition probabilities and mean dwelling time, BI-M provides a reasonable estimate. Actually, even in the most challenging case of Muller's potential with $\Gamma=10$ and $24$ anchors, the largest difference in transition probabilities between LPT-M and BI-M remains under 0.1. In the example of deca-alanine unfolding, the largest difference in transition probabilities between LPT-M and BI-M is only about $0.06$. The difference in mean dwelling time is even smaller. These subtle differences likely stem from the BI-M formalism, wherein the estimation of one probability ratio $K_{\alpha\beta}$ relies on the assessment of another probability ratio $P(FHP|x)$. This estimation process is more intricate than the more direct approach of LPT-M. The differences in transition probabilities are ultimately magnified when solving the Eqs. \eqref{MFPT absorb} or \eqref{flux cyc} for MFPT. 

\begin{table}
\centering
\caption{The number of force builds per milestone ($N_{\mathrm{FB}}$) of CM, LPT-M, BI-M and a long trajectory simulation with cyclic boundary condition (Long-C) in deca-alanine unfolding in vacuum.}\label{Nfb unfolding}
\begin{threeparttable}
\begin{tabular}{cccccc}\toprule
    &\quad  CM\quad &\quad LPT-M\quad &\quad BI-M\quad &\quad Long-C\quad\\
\midrule
{$N_{\mathrm{FB}}$ $(\times 10^6)$}  &\quad   $1.1$ \quad &\quad $1.5$ \quad& \quad $3.6$\quad &\quad $235$\quad\\
\bottomrule
\end{tabular}
\end{threeparttable}
\end{table}

\section{Conclusion}\label{Conclusion}
MFPT is the key kinetic output of Milestoning, whose accuracy crucially depends on FHPD. 
Non-equilibrium stationary FHPD of the partial path ensemble $A\rightarrow B$ is required for exact MFPT calculations. This has been extensively discussed in TPT\cite{TPT06,TPT10}, NEUS\cite{NEUS09}, FFS\cite{FFlux06,FFlux09}, and trajectory tilting\cite{Tilting09} approach. However, constructing an accurate approximation of the non-equilibrium stationary FHPD can be expensive, as it requires tracing trajectories backwards to either the reactant $A$ or the product $B$. To address this, we develop two algorithms, LPT-M and BI-M, to approximate equilibrium FHPD. 
While the calculated MFPT from these methods are not exact, both LPT-M and BI-M improve significantly over the conventional CM method.

LPT-M is particularly preferable as its high accuracy approaching the accuracy limit of an equilibrium long trajectory simulation, better robustness with respect to the number of intermediate milestones, and only a modest increase (about $50\%$) in computational costs compared to CM.
Notably, LPT-M's wall-clock time is actually the same as CM, since the forward and backward trajectories in LPT-M can run in parallel.

Furthermore, we introduce a novel ratio function $R_{A\rightarrow B}$ that quantifies the contribution of path ensemble $A\rightarrow B$ to equilibrium FHPD on each milestone. 
Leveraging this insight, we develop a local iteration strategy for exact MFPT calculation on milestones with low $R_{A\rightarrow B}$. Based on LPT-M/BI-M, this approach exhibits lower computational costs and better stability compared to the existing ExM method.

\begin{acknowledgments}
This work was supported by the Qilu Young Scholars Program of Shandong University.
\end{acknowledgments}

\section*{Data Availability Statement}
The data that support the findings of this study are available within the article and its supplementary material.

\section*{Conflicts of interest}
There are no conflicts to declare.

\section*{Supporting Information}
Anchor positions on Mueller's potential in Fig. 3, milestone corresponding list in Fig. 4, MFPT values adapted from Fig. 5, MFPT values and number of force builds per milestone adapted from Fig. 6, the error plot of transition probabilities and mean dwelling time, the ratio function $R_{A\rightarrow B}$ profile along the reaction pathway in Mueller's potential and deca-alanine unfolding process.

\clearpage
\newpage

\appendix

\clearpage
\newpage

\bibliographystyle{achemso1}
\bibliography{FHPD_abbrev}

\newpage
TOC
\begin{figure}[h]
\centering
\begin{tabular}{cc}
\includegraphics[height=7cm]{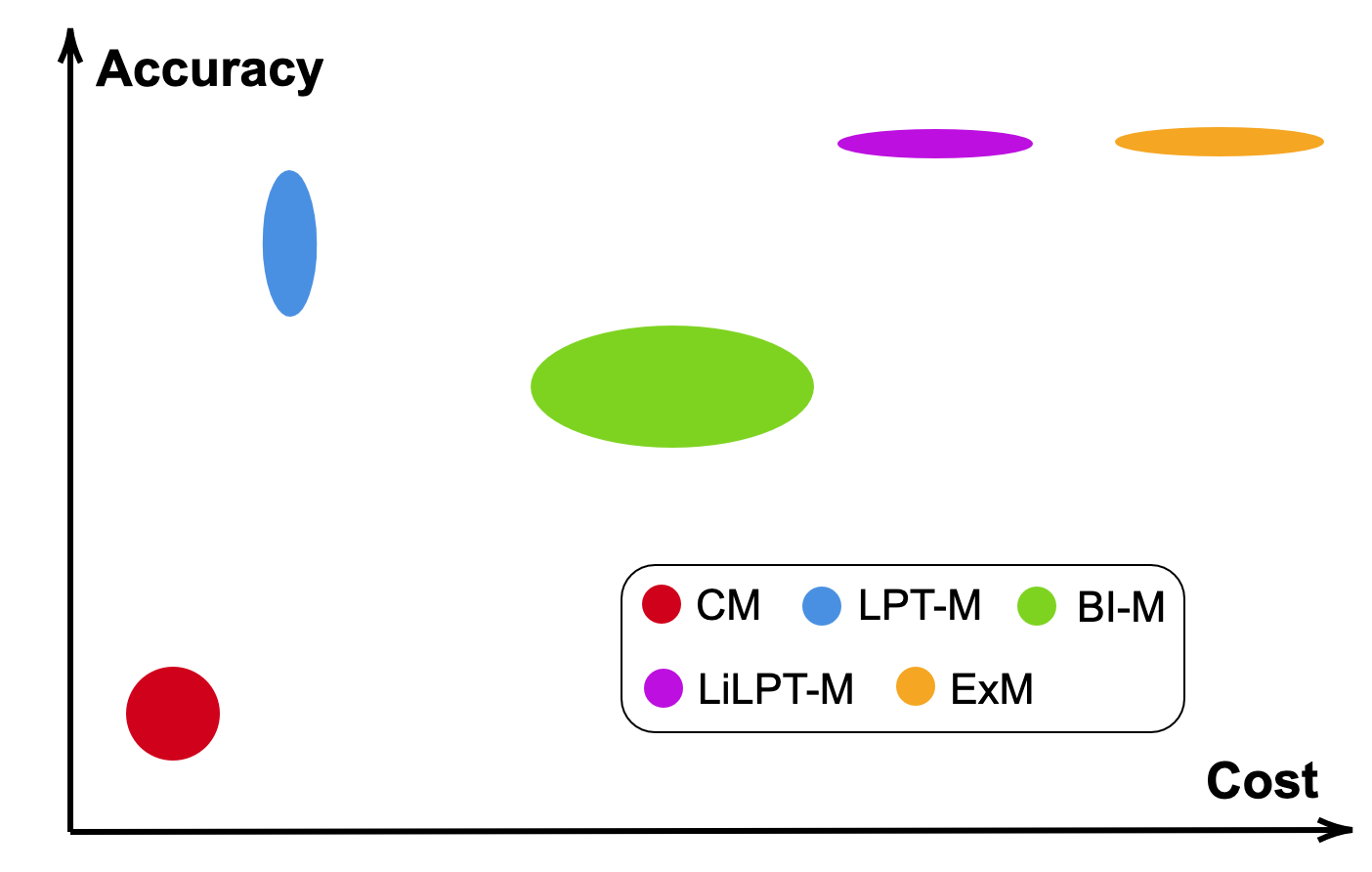}
\end{tabular}
\end{figure}
\end{document}